\documentclass[a4paper,11pt]{article}
\pdfoutput=1 

\usepackage{jheppub} 

\usepackage[T1]{fontenc} 
\usepackage[utf8]{inputenc}
\usepackage{amsmath,amssymb,amstext,amsfonts} 

\usepackage[dvipsnames]{xcolor}
\usepackage{graphicx}
\usepackage{hyperref}
\usepackage{cancel} 
\usepackage{float}
\usepackage{bbm}
\usepackage{mathtools}

\usepackage{lscape}
\usepackage{siunitx}
\usepackage{subcaption}

\usepackage{cleveref}
\makeatletter
\renewcommand*\env@matrix[1][\arraystretch]{%
  \edef\arraystretch{#1}%
  \hskip -\arraycolsep
  \let\@ifnextchar\new@ifnextchar
  \array{*\c@MaxMatrixCols c}}
\makeatother

\makeatletter
\g@addto@macro\bfseries{\boldmath}
\makeatother

\newcommand{\mr}{m_r}
\newcommand{\psir}{\psi_r}

\newcommand{\eps}{\epsilon}
\newcommand{\myperiod}{\varpi}
\newcommand{\qbar}{q}

\DeclareMathOperator{\K}{K}
\DeclareMathOperator{\rd}{d}

\def\beq{\begin{equation}}
\def\eeq{\end{equation}}
\def\bsp#1\esp{\begin{split}#1\end{split}}

\newenvironment{gleichung*}{\begin{equation*}\begin{aligned}}{\end{aligned}\end{equation*}}
\usepackage{slashed}

\usepackage{tikz}
    \usetikzlibrary{positioning}
    \usetikzlibrary{arrows}
    \usetikzlibrary{shapes}
    \usepackage{pgfplots}
    \usetikzlibrary{calc}
    \usetikzlibrary{decorations.markings}
     \usetikzlibrary{decorations.pathmorphing}
    \tikzset{snake it/.style={decorate, decoration=snake}}
\def\centerarc[#1](#2)(#3:#4:#5) 
    { \draw[#1] ($(#2)+({#5*cos(#3)},{#5*sin(#3)})$) arc (#3:#4:#5); }
    \usepackage{booktabs}

\definecolor{cnblue}{RGB}{7,82,154}

\title{\boldmath 
On the  electron self-energy to three loops in QED}

\author[a]{Claude Duhr,} 
\author[b]{Federico Gasparotto,}
\author[c]{Christoph Nega,} 
\author[c]{Lorenzo Tancredi}
\author[b]{and Stefan Weinzierl} 

\affiliation[a]{Bethe Center for Theoretical Physics, Universit\"at Bonn, D-53115, Germany}
\affiliation[b]{PRISMA Cluster of Excellence, Institut für Physik, Johannes Gutenberg-Universität Mainz,
D-55099 Mainz, Germany}
\affiliation[c]{Technical University of Munich, TUM School of Natural Sciences, Physics Department, James-Franck-Straße 1, D-85748 Garching, Germany}

\emailAdd{cduhr@uni-bonn.de}
\emailAdd{fgasparo@uni-mainz.de}
\emailAdd{c.nega@tum.de}
\emailAdd{lorenzo.tancredi@tum.de}
\emailAdd{weinzierl@uni-mainz.de}

\preprint{\begin{minipage}[t]{8cm}\begin{flushright} 
BONN-TH-2024-12 \\
MITP/24-065 \\
TUM-HEP-1518/24
      \end{flushright}\end{minipage}}

\abstract{

We compute the electron self-energy in Quantum Electrodynamics to three loops
in terms of iterated integrals over kernels of elliptic type. We make use of the differential equations
method, augmented by an $\epsilon$-factorized basis, which allows us to gain full control over the differential forms
appearing in the iterated integrals to all orders in the dimensional regulator.
We obtain compact analytic expressions, for which we provide generalized series expansion
representations that allow us to evaluate the result numerically for all values of the
electron momentum squared. As a by-product, we also obtain $\epsilon$-resummed results for the
self-energy in the on-shell limit $p^2 = m^2$, which we use to recompute the known
three-loop renormalization constants in the on-shell scheme.
}

\begin{document} 
\maketitle
\flushbottom


\section{Introduction}
\label{sec:intro}

The electron self-energy is one of the fundamental quantities of interest in Quantum Electrodynamics (QED).
Its calculation in perturbation theory, and the interpretation of the corresponding result, 
has attracted a lot of attention from the first days of quantum field theory (QFT), working also as a motivation
to lay down the foundations towards a general theory of renormalization.
Indeed, as of today the computation of the one-loop corrections to the electron self-energy in QED 
is an integral part of most standard QFT courses as one of the
first examples of applications of QFT perturbation theory beyond the leading order.

From the first calculations by Schwinger and Feynman (see ref.~\cite{Bovy:2006dr} for a summary of these historical developments), a lot has changed in the way we address the calculation of loop Feynman integrals.
In fact, the increasing precision of high- and low-energy particle experiments on the one hand, and on the other
the formal interest in searching for general mathematical patterns in perturbative QFT, have triggered
the development of new mathematical methods which have enormously simplified these calculations.
These include integration-by-parts identities~\cite{Tkachov:1981wb,Chetyrkin:1981qh} and the Laporta algorithm~\cite{Laporta:2004rb},
the differential equation method~\cite{Kotikov:1990kg,Remiddi:1997ny,Gehrmann:1999as},
and a new understanding of the theory of the special functions involved, in particular multiple polylogarithms~\cite{Kummer,Remiddi:1999ew,Goncharov:1995,Goncharov:1998kja,Vollinga:2004sn,Duhr:2011zq},
and more general iterated integrals~\cite{ChenSymbol} over logarithmic differential forms (dlog forms).
As a result, extremely sophisticated calculations have become possible, including scattering amplitudes
for processes involving three, four or five external particles, up to four, three and two loops, respectively, see refs.~\cite{FebresCordero:2022psq,Caola:2022ayt} and
references therein. 

Despite these impressive results, up until recently a fully analytic result for the two-loop corrections
to the very fundamental (and apparently simpler) electron self-energy were not available. 
The first attempt to this calculation can be traced back to A. Sabry~\cite{Sabry} in 1962,
who showed that the discontinuity of the self-energy contains new mathematical objects which cannot
be expressed as iterated integrals over dlog forms, namely elliptic integrals. While Sabry
could provide series representations for the two-loop self-energy, we had to wait more than
half a century for its first fully analytic calculation~\cite{Honemann:2018mrb}. This became possible
thanks to recent advances in our understanding of iterated integrals defined over more general classes
of differential forms, in particular those defined over genus-one curves. 
If the genus-one curve has in addition one marked point, the curve is also known as an elliptic curve~\cite{MR2024529}.
Interestingly, while these new geometries can give rise to extremely rich mathematical structures,
the two-loop electron self-energy turned out to be expressible in terms of a very well-defined
subclass of these integrals, whose differential forms have especially simple transformation properties
under a subgroup of modular transformations, namely iterated integrals
of modular forms for the congruence subgroup $\Gamma_1(6)$~\cite{Adams:2017ejb}.

In turn, standard iterated integrals over modular forms are well known not to be sufficient to describe elliptic 
scattering amplitudes with non-trivial kinematic dependence already at two loops, and it remains an 
open question if a well-defined extension of these integrals can accommodate a large enough
number of physically relevant problems. An especially powerful technique to investigate the space of
functions characterizing a given physical problem is the differential equation's method, in particular
when augmented by the idea of a canonical basis~\cite{ArkaniHamed:2010gh,Henn:2013pwa}.
These are bases of master integrals which fulfill especially simple differential equations,
where the dependence on the dimensional regulator $\epsilon$ is entirely factorized from the dependence
on the kinematics. In addition, if the
differential forms appearing in the connection matrix are dlog forms, the corresponding
solutions can be typically expressed as iterated integrals of polylogarithmic type. Though care is needed, because there may be exceptions, where an iterated integral of dlog forms cannot be expressed in terms of polylogarithms~\cite{Duhr:2020gdd}. While the notion of a canonical basis remained for long time confined to dlog forms,
recently important steps forward have been made to understand its generalization
to elliptic and Calabi-Yau geometries~\cite{Adams:2016xah,Adams:2018bsn,Pogel:2022ken,Frellesvig:2021hkr,Dlapa:2022wdu,Frellesvig:2023iwr,Gorges:2023zgv,Pogel:2022yat,Behring:2023rlq}. A basis in this form allows one  not only to make general considerations about the mathematical structure of the solutions 
to all orders in the dimensional regulator, but also provides a way to investigate 
the analytic properties of the solution  close to the physical number of space-time dimensions $D=4$.

Armed with these new mathematical tools, we are now in the position to push the study of more fundamental
quantities in QFT to higher orders in perturbation theory.
In this paper, we focus on the three-loop corrections to the electron self-energy
in QED, retaining full dependence on the electron mass.
We will see that, while the analytic structure of the solution
becomes much richer when going from two to three loops, the very same elliptic curve
encountered at two loops remains at the core of the calculation. This 
quantity constitutes a perfect laboratory
to investigate which types of iterated integrals are relevant
in relativistic quantum field theories involving elliptic curves.

The rest of this paper is organized as follows. In section~\ref{sec:def} we provide general definitions 
and conventions, including the families of Feynman integrals that we use for our calculation. 
We continue in section~\ref{sec:epsbasis}, providing an $\epsilon$-factorized basis for the problem at hand
and presenting the relevant differential forms.
We solve the differential equations in terms of iterated integrals in section~\ref{sec:iterint} and show
that some differential forms do not appear in the final physical result in $D=4$ space-time dimensions.
In particular, we show that with our basis the poles contain the same class of functions already encountered at two loops, as expected
from the general theory of UV renormalization. We also study in detail the structure of the solution
close to the on-shell limit. We obtain fully $\epsilon$-resummed solutions for the divergent
branches, which allow us to easily rederive all on-shell renormalization conditions without
having to resort to a separate calculation, see section~\ref{sec:local1}. From the explicit analytic results, we
see that the bare three-loop self-energy develops a pole as 
$p^2 \to m^2$, which is removed once
the mass is properly renormalized in the on-shell scheme. We describe renormalization in general in 
section~\ref{sec:ren}. We also provide analytic solutions close to the singular point $p^2=0$ in section~\ref{sec:local0}.
For the explicit numerical evaluation of the results, we provide fast converging series expansions valid in each
analytic region in section~\ref{sec:global}. There, we also investigate
the effect of employing so-called Bernoulli-like variables~\cite{tHooft:1978jhc} 
to accelerate the convergence of the corresponding series.
Finally, we draw our conclusions in section~\ref{sec:conc}.


\section{Definitions and Feynman diagrams calculation}
\label{sec:def}

We consider the QED corrections to the electron self-energy $\hat{\Sigma}(p,m)$, 
with one single massive  fermion of mass $m$. Note that throughout the paper, we use the symbol $m$ 
to denote the bare mass.
The momentum of the off-shell electron is indicated by $p$. For later convenience, we also introduce
the massless ratio
\begin{equation}
    x = \frac{p^2}{m^2}\,, \label{eq:xdef}
\end{equation}
in terms of which we can fully describe the analytic structure of the self-energy. We normalize the self-energy such that,
after resumming all one-particle irreducible diagrams, the bare electron propagator
can be written as
\begin{align}
    \Pi_F(p) &= \frac{i}{\slashed{p} - m +\hat{\Sigma}(p,m )} \,. \label{eq:bareProp}
\end{align}  
We work in a generic $R_\xi$ gauge,
where the photon propagator reads
\begin{equation}
\Pi_\gamma^{\mu \nu}(q) = - \frac{i}{q^2 } \left[ g^{\mu \nu} - (1-\xi) \frac{q^\mu q^\nu}{q^2} \right]\,.
\label{eq:propphoton}
\end{equation}

As customary, we decompose the self-energy into two independent spinor structures,
\begin{equation}
\hat{\Sigma}(p,m) = \Sigma_V(p,m) \slashed{p}  + \Sigma_S (p,m)\, m \mathbbm{1} \, ,
\label{eq:defSigma}
\end{equation}
where the two scalar functions $\Sigma_V(p,m)$ and $\Sigma_S(p,m)$ can be evaluated from
a Feynman diagram representation of the self-energy, as
\begin{equation}
\Sigma_V(p,m) = \frac{1}{4 p^2} {\rm Tr} \left[ \slashed{p} \,\hat{\Sigma}(p,m)  \right]\,, \qquad 
\Sigma_S(p,m) = \frac{1}{4 m} {\rm Tr} \left[ \hat{\Sigma}(p,m) \right]\,. \label{eq:defSigmaVS}
\end{equation}
Here we used the common convention $ {\rm Tr}[ \mathbbm{1} ] = 4$ in dimensional regularization in $D=4-2\epsilon$ dimensions.

The scalar functions so introduced can be expanded in a perturbative
series in the electric coupling $\alpha$, where $e = \sqrt{4 \pi \alpha}$ is the bare 
(positive) electron charge,
\begin{align}
\Sigma_N(p,m) =
 \sum_{\ell=0}^\infty \left(  \frac{\alpha}{ \pi} C(\epsilon)\right)^\ell 
\Sigma_N^{(\ell)}(p,m) \,,
\end{align}
where  $N=V,S$ and we introduced the $\epsilon$-dependent normalization
\begin{equation}
    C(\epsilon) =  \Gamma(1+\epsilon) (4 \pi)^\epsilon (m^2)^{-\epsilon}\,. \label{eq:defCeps}
\end{equation}
As usual, $\Sigma_N^{(\ell)}(p,m)$ receives contributions from
Feynman diagrams at $\ell$ loops.

\begin{figure}[h]
\begin{center}
	\includegraphics[width=13cm]{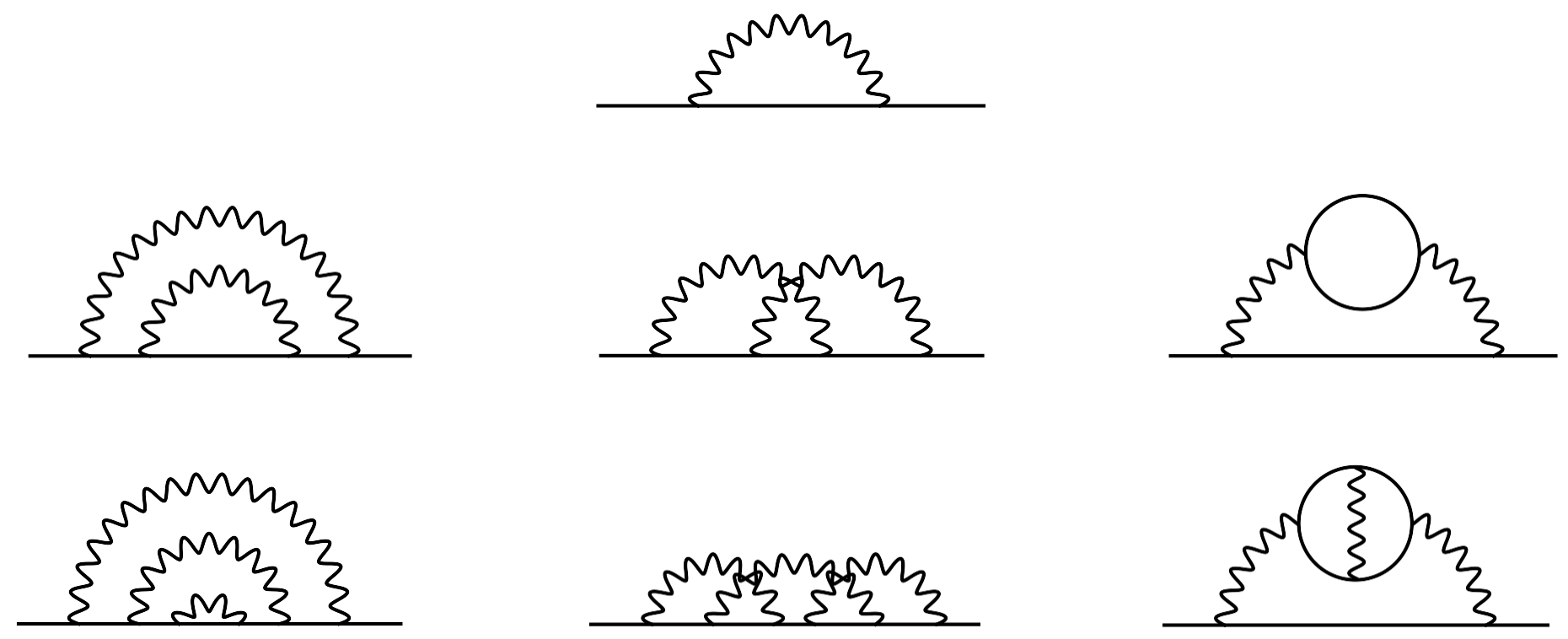}
\caption{Some of the Feynman diagrams, necessary for the electron self-energy up to three loops. In total, there are 1, 3 and 20 diagrams at one, two and three loops, respectively.}
\label{fig:diagrams}
\end{center}
\end{figure}

To generate the relevant Feynman diagrams up to three loops, we use \texttt{QGRAF}~\cite{Nogueira:1991ex} and \texttt{FeynArts}~\cite{Hahn:2000kx}.
We find  1 diagram at one loop, 3 diagrams at two loops, and 20 diagrams at three loops, see fig.~\ref{fig:diagrams} for some explicit examples.
We perform all relevant spinor algebra to extract the scalar functions in eq.~\eqref{eq:defSigmaVS} 
with \texttt{FORM}~\cite{Vermaseren:2000nd,Kuipers:2012rf,Ruijl:2017dtg} and \texttt{FeynCalc}~\cite{Mertig:1990an,Shtabovenko:2016sxi,Shtabovenko:2020gxv}. 

Focusing on the three-loop case, 
we find that all integrals can be mapped in terms of two integrals families
\begin{equation}
I^\text{fam}_{n_1 \, n_2 \, n_3 \, n_4 \, n_5 \, n_6 \, n_7 \, n_8 \, n_9}
= e^{3 \epsilon \gamma} \int \prod_{\ell=1}^3 \frac{\mathrm d^D k_\ell}{i \pi^{D/2}} \frac{1}{D_1^{n_1} \hdots D_9^{n_9}}
\end{equation}
where $\gamma$ is the Euler Mascheroni constant and $\text{fam} =A,B$.
The propagators for the two families are defined in table~\ref{tab:props}.

\begin{table}[H]
\begin{center}
\begin{tabular}{| m{2.5cm} || m{3.5cm} | m{3.5cm} |} 
 \hline
 Denominator & Integral family A & Integral family B  \\[5pt]
 \hline
$D_1$ & $k_1^2$                  & $k_1^2$       \\
$D_2$ & $k_2^2$                  & $k_2^2$  \\
$D_3$ & $k_3^2 $                 & $k_3^2 $   \\
$D_4$ & $(k_1-p)^2 -m^2$         & $(k_1-p)^2 -m^2$ \\
$D_5$ & $(k_2-p)^2 -m^2 $        & $(k_2-p)^2 -m^2 $ \\
$D_6$ & $(k_3-p)^2 -m^2 $        & $(k_3-p)^2 -m^2 $    \\
$D_7$ & $(k_1+k_3-p)^2 -m^2$     & $(k_1-k_2)^2 $   \\
$D_8$ & $(k_2+k_3-p)^2 -m^2$     & $(k_1+k_3)^2 -m^2$ \\
$D_9$ & $(k_1+k_2+k_3-p)^2 -m^2$ & $(k_1-k_2+k_3)^2 -m^2$ \\
 \hline
\end{tabular}  \caption{\label{tab:props}Definitions of the propagators of the two scalar integral families $A, B$.}
\end{center}
\end{table}
It turns out that, after using symmetry identities and integration-by-parts (IBP) identities, 
family 
$B$ does not introduce any additional master integral that cannot
be expressed in terms of family $A$. We therefore ignore it in what follows.
We reduce all scalar integrals to master integrals using \texttt{Reduze 2}~\cite{Studerus:2009ye,vonManteuffel:2012np},  \texttt{Kira 2}~\cite{Maierhofer:2017gsa,Klappert:2019emp,Klappert:2020nbg} and also partly \texttt{LiteRed}~\cite{Lee:2012cn}
using IBP identities
and we find a $51$-dimensional basis, as quoted above, all from family $A$:
\allowdisplaybreaks
\begin{align}
&I^A_{000111000} \,, \qquad 
&I^A_{100111000} \,, \qquad 
&I^A_{101010100} \,, \qquad 
&I^A_{101010200} \,, \qquad 
&I^A_{000111100} \,, \nonumber \\ 
&I^A_{000111200} \,, \qquad 
&I^A_{110000110} \,, \qquad 
&I^A_{010100110} \,, \qquad
&I^A_{010100120} \,, \qquad 
&I^A_{020100110} \,, \nonumber \\ 
&I^A_{000110110} \,, \qquad
&I^A_{111000001} \,, \qquad 
&I^A_{111000002} \,, \qquad 
&I^A_{110111000} \,, \qquad 
&I^A_{111010100} \,, \nonumber \\
&I^A_{111010200} \,, \qquad 
&I^A_{010111100} \,, \qquad
&I^A_{010111200} \,, \qquad 
&I^A_{111000110} \,, \qquad 
&I^A_{111000120} \,, \nonumber \\
&I^A_{100110110} \,, \qquad 
&I^A_{000111110} \,, \qquad 
&I^A_{000111120} \,, \qquad
&I^A_{000112110} \,, \qquad 
&I^A_{111100001} \,, \nonumber \\
&I^A_{101010101} \,, \qquad 
&I^A_{101010102} \,, \qquad 
&I^A_{101010201} \,, \qquad 
&I^A_{111111000} \,, \qquad
&I^A_{101111100} \,, \nonumber \\
&I^A_{111100110} \,, \qquad 
&I^A_{111100120} \,, \qquad 
&I^A_{110110110} \,, \qquad 
&I^A_{110110120} \,, \qquad 
&I^A_{100111110} \,, \nonumber \\
&I^A_{111110001} \,, \qquad 
&I^A_{111110002} \,, \qquad 
&I^A_{111120001} \,, \qquad 
&I^A_{101110101} \,, \qquad 
&I^A_{111000111} \,, \nonumber \\
&I^A_{111000112} \,, \qquad
&I^A_{011100111} \,, \qquad 
&I^A_{011100121} \,, \qquad 
&I^A_{111111100} \,, \qquad 
&I^A_{110111110} \,, \nonumber \\
&I^A_{111110101} \,, \qquad 
&I^A_{111110102} \,, \qquad
&I^A_{111100111} \,, \qquad 
&I^A_{111111110} \,, \qquad 
&I^A_{111110111} \,, \nonumber \\
&I^A_{111110112} \,. & \label{eq:misI}
\end{align}

\begin{figure}[h]
\begin{center}
	\includegraphics[width=0.8\textwidth]{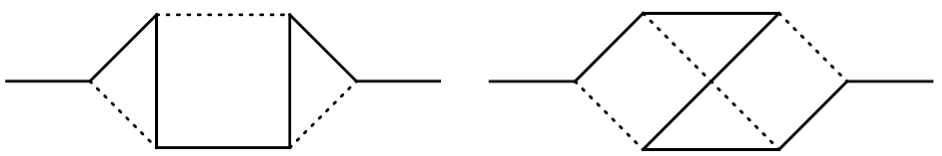}
\caption{Planar and non-planar top sector diagrams relevant for the calculation of the three-loop electron propagator. From the graphs, one can see that there can never be a cut through four massive particles, which excludes contributions proportional to the three-loop massive banana graph.}
\label{fig:topsectors}
\end{center}
\end{figure}

As one can easily see from the top sectors
drawn in fig.~\ref{fig:topsectors}, fermion number conservation
in QED implies that one can never produce diagrams 
with a four massive particle cut up to three loops.
The latter is related to the K3 geometry 
associated to the three-loop massive banana graph~\cite{Primo:2017ipr,Broedel:2019kmn,Bonisch:2021yfw,Pogel:2022yat}.
In fact, as one can easily demonstrate by direct calculation,
it turns out that all master integrals have leading singularities that correspond either to trivial dlog-forms,
or at most to the same elliptic curve of the two-loop sunrise graph, which fully characterizes the geometry
of the electron propagator one order lower in perturbation theory. 

\subsection{The elliptic curve}

\begin{figure}[h!]
    \centering
    \includegraphics[width=4cm]{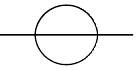}
    \caption{The two loop sunrise  graph, whose maximal
    cut is associated to the elliptic curve defined by eq.~(\ref{eq:curve}).}
    \label{fig:2L_elliptic_sunrise}
\end{figure}

While the elliptic curve appearing in the sunrise graph has already been studied at full length in the literature~\cite{Sabry,Broadhurst:1993mw,Laporta:2004rb,MullerStach:2011ru,Bloch:2013tra,Remiddi:2013joa},
we recall here some of its properties for convenience of the reader and to establish our notations.
The elliptic curve associated to the maximal cut of the two-loop sunrise 
(see fig.~\ref{fig:2L_elliptic_sunrise}) can be defined 
in terms of the kinematical invariants of the propagator by
the following fourth-order equation
\begin{equation}
    Y^2 = P_4(X) = X\big(X-4m^2\big)\big(X-(\sqrt{p^2}-m)^2\big)\big(X-(\sqrt{p^2}+m)^2\big)\,. \label{eq:curve}
\end{equation}

The periods of the elliptic curve satisfy
the following second-order Picard-Fuchs equation
\begin{equation}
\left[ \left( x \frac{\mathrm d}{\mathrm dx} \right)^2
+  \left(\frac{1}{x-1}+\frac{9}{x-9}+2\right)\left( x \frac{\mathrm d}{\mathrm dx}\right)
+\frac{27}{4 (x-9)}+\frac{1}{4 (x-1)}+1 \right] \varpi(x) = 0\,, \label{eq:pfell}
\end{equation}
where we have used the dimensionless ratio $x$ defined in eq.~\eqref{eq:xdef}.
It is well known that close to any regular singular point, this equation admits two solutions, 
a regular one and one which diverges logarithmically. 
In a neighborhood of a regular-singular point, we always denote the regular or \emph{holomorphic} solution by $\varpi_0(x)$.
Note that a solution, which is holomorphic at one regular singular point, need not be holomorphic at another regular-singular point (see also refs.~\cite{Duhr:2019rrs,Frellesvig:2023iwr}).
If we discuss different regular singular points $p$ at the same time, we will use superscripts like $\varpi_0^{[p]}(x)$ to indicate the solution that is holomorphic in  a neighborhood of the point $p$. 

In this specific case, there are four regular singular points $x\in \{0,1,9,\infty \}$.
For convenience, we provide here our choice of the solutions close to each singular point, in terms of
their series expansion through the first four orders. Our basis is chosen such that the branch cut of the logarithmic solution is on the negative real axis of the local variable.

\begin{itemize}
    \item For $x=0^+$, we define the two solutions as
\begin{equation}
\begin{aligned}
  \varpi_0^{[0]}(x) &= 1+\frac{x}{3}+\frac{5 x^2}{27}+\frac{31 x^3}{243}+\frac{71 x^4}{729} + \mathcal O(x^5) \,, \\
  \varpi_1^{[0]}(x) &= \varpi_0^{[0]}(x)\log(x) + \frac{4 x}{9}+\frac{26 x^2}{81}+\frac{526 x^3}{2187}+\frac{1253 x^4}{6561} + \mathcal O(x^5)\,.
  \label{eq:defper0}
\end{aligned}
\end{equation}

\item Close to $x=1^+$, it is convenient to use the local variable $u=x-1$, in terms of which the two solutions are given by 
\begin{equation}
\begin{aligned}
  \varpi_0^{[1]}(u) &= 1-\frac{u}{4}+\frac{5 u^2}{32}-\frac{7 u^3}{64}+\frac{173 u^4}{2048} + \mathcal O(u^5)\,, \\
  \varpi_1^{[1]}(u) &= \varpi_0^{[1]}(u) \log(u)  -\frac{3 u}{8}+\frac{33 u^2}{128}-\frac{25 u^3}{128}+\frac{2561 u^4}{16384} + \mathcal O(u^5)\,.
  \label{eq:defper1}
\end{aligned}
\end{equation}

\item Close to $x=9^+$, we introduce the local variable $v=x-9$, and find for the solutions
\begin{equation}
\begin{aligned}
  \varpi_0^{[9]}(v) &=  1-\frac{v}{12}+\frac{7 v^2}{864}-\frac{13 v^3}{15552}+\frac{133 v^4}{1492992} + \mathcal O(v^5)\,, \\
  \varpi_1^{[9]}(v) &= \varpi_0^{[9]}(v) \log(v)   -\frac{5 v}{72} +\frac{97 v^2}{10368}-\frac{157 v^3}{139968}+\frac{14081 v^4}{107495424} + \mathcal O(v^5)\,. \label{eq:defper9}
\end{aligned}
\end{equation}

\item Finally, at $x=+\infty$, we use the local variable $w=1/x$, and define the two solutions as
\begin{equation}
\begin{aligned}
  \varpi_0^{[\infty]}(w) &=  w+3 w^2+15 w^3+93 w^4 + \mathcal O(w^5)\,, \\
  \varpi_1^{[\infty]}(w) &= \varpi_0^{[\infty]}(w) \log(w)   +4 w^2+26 w^3+\frac{526 w^4}{3} + \mathcal O(w^5)\,. \label{eq:defperinf}
\end{aligned}
\end{equation}
\end{itemize}
The expressions above are solutions of the second-order differential equation~\eqref{eq:xdef}
obtained with the Frobenius method, and they are sufficient for our purposes.

Let us briefly discuss the relation of the Frobenius solutions to the integral periods of the elliptic curve.
To make this connection, we first denote the roots
of the quartic polynomial by (we set $m=1$ and use the dimensional variable $x$)
\begin{equation}
\begin{aligned}
    X_1 & = 4,
     &
    X_2 & = \left(1+\sqrt{x}\right)^2,
    &
    X_3 & = \left(1-\sqrt{x}\right)^2,
    &
    X_4 & = 0
\end{aligned}
\end{equation}
and set
\begin{equation}
        Z_1  =  \left(X_3-X_2\right)\left(X_4-X_1\right), \quad
        Z_2  =  \left(X_2-X_1\right)\left(X_4-X_3\right), \quad
        Z_3  =  \left(X_3-X_1\right)\left(X_4-X_2\right).
\end{equation}
The modulus squared $k^2$ of the elliptic curve is given by
\begin{equation}
    k^2 = \frac{Z_1}{Z_3}.
\end{equation}
The elliptic curve has one holomorphic one-form, which is defined up to a prefactor.
The conventional choice is $\tfrac{\rd\! X}{Y}$, and we will adopt this choice.
The periods of the elliptic curve are then the integrals of the holomorphic one-form
along the cycles of the curve.
Choosing two independent cycles, we find that for one particular choice the two independent periods are given by
\begin{equation}\label{eq:pers}
    \begin{aligned}
        \varpi_0^{[\mathrm{per}]} & = \frac{4}{\sqrt{Z_3}} \K\left(k^2\right),
         &
         \varpi_1^{[\mathrm{per}]} &= \frac{4i}{\sqrt{Z_3}} \K\left(1-k^2\right),
    \end{aligned}
\end{equation}
where $\K(\lambda)$ denotes the complete elliptic integral of the first kind,
\beq
\K(\lambda) = \int_0^1\frac{\rd x}{\sqrt{(1-x^2)(1-\lambda x^2)}}\,.
\eeq
All other choices for the two independent cycles are related by a modular
$\mathrm{SL}_2({\mathbb Z})$-transfor\-ma\-tion to the choice above.
One defines the modular parameter $\tau$ and the nome squared $\qbar$ by
\begin{equation}
    \begin{aligned}
        \tau & = \frac{\varpi_1^{[\mathrm{per}]}}{\varpi_0^{[\mathrm{per}]}},
        &\qquad
        \qbar & = e^{2\pi i \tau}. \label{eq:tau}
    \end{aligned}
\end{equation}
The relation between the periods of the elliptic curve and the Frobenius solutions around the regular singular point $x=0$ is given by
\begin{equation}
\begin{aligned}
    \varpi_0^{[\mathrm{per}]} & = \frac{2}{3}\sqrt{3} \pi \varpi_0^{[0]},
    &
    \varpi_1^{[\mathrm{per}]} & = \frac{2}{3}\sqrt{3} \pi \frac{1}{2\pi i}\left( \varpi_1^{[0]} - 2 \ln(3) \varpi_0^{[0]} \right).
\end{aligned}
\end{equation}
Similar relations can be worked out around all other regular singular points.
We see that the Frobenius solutions are ${\mathbb C}$-linear combinations 
of the periods of the elliptic curve, but not ${\mathbb Z}$-linear combinations.
We stress that the Frobenius solutions defined in~\cref{eq:defper0,eq:defper1,eq:defper9,eq:defperinf} are sufficient for our purpose, and in the 
rest of the paper we will only use those.


\section{Differential equations in $\epsilon$-form}
\label{sec:epsbasis}

As the first step towards obtaining analytic results for the electron self-energy,
we start by deriving a system of differential equations for the master integrals
identified in the previous section. By using IBP identities,
it is easy to show that the integrals fulfill a system of linear differential equations
with rational coefficients~\cite{Kotikov:1990kg,Remiddi:1997ny}. 
For simplicity, we use the dimensionless ratio $x = p^2/m^2$ defined in eq.~\eqref{eq:xdef}
and introduce the vector $\vec{I}$, which contains the master integrals given in eq.~\eqref{eq:misI}.
The differential equations can then be written in matrix notation as 
\begin{equation}
    \frac{\mathrm d}{\mathrm dx} \vec{I} = B(\epsilon, x) \vec{I} \,,
\end{equation}
where $B(\epsilon,x)$ is a matrix of rational functions in $x$ and in the dimensional 
regulator $\epsilon$.
The equations are naturally in block-triangular form, since integrals with  fewer
propagators can only appear as inhomogeneous terms of integrals which contain 
those propagators as subsets of theirs. 

While we could attempt to solve the differential equations directly in this form, it is extremely
convenient to perform a rotation in the space of master integrals and define a new
basis of master integrals that fulfills a so-called $\epsilon$-factorized system of 
differential equations~\cite{Henn:2013pwa}.
More explicitly, we define a new basis of integrals $\vec{J}$  through
\begin{equation}
    \vec{J} = A(\epsilon,x) \vec{I}\,,
\end{equation}
and we would like to find a matrix $A(\epsilon,x)$ such that the new basis of master
integrals fulfills
\begin{align}
    \frac{\mathrm d}{\mathrm dx} \vec{J} = \epsilon \, G(x) \vec{J}\,,
    \qquad \epsilon\, G(x) = A \, B\, A^{-1} + \frac{\mathrm dA}{\mathrm dx} A^{-1}\,.\label{eq:canf}
\end{align}
In this form, the differential equations can easily be solved as series expansion in
$\epsilon$. The matrix $G(x)$ can in general 
be written as follows
\begin{equation}
    G(x) = \sum_i G_i\, f_i(x)\,,\label{eq:GMf}
\end{equation}
where $G_i$ are numerical matrices and the $f_i(x)$ are functions of the kinematical
variable $x$. 
Importantly, the functions $f_i(x)$ determine the analytic structure
of the solutions to all orders in $\epsilon$. Using the language
of differential forms we can write 
\begin{equation}
f_i(x) \mathrm dx = \omega_i\,,
\end{equation}
such that the system of differential equations takes the form
\begin{equation}
    \mathrm d\vec{J} = \epsilon \left( \sum_i G_i\, \omega_i \right) \vec{J}\,. \label{eq:diffeps}
\end{equation}
Its solutions can formally be written as a path-ordered exponential,
\begin{equation}
    \vec{J}(x) = \mathbb{P}\exp{ \left[ \epsilon \sum_i G_i \int_\gamma  \omega_i \right] } \vec{J}_0 \,,
\end{equation}
where $\mathbb{P}$ is the path-ordering operator,  
$\vec{J}_0$ is the boundary condition at $x=x_0$ and $\gamma$ is a path
that connects the points $x_0$ to the generic point $x$.
In this form, it becomes obvious that at all orders in $\epsilon$
the integrals $\vec{J}$ can be written as linear combinations of (Chen)
iterated integrals~\cite{ChenSymbol} over the forms $\omega_i$.
In the problem under study we only deal with one kinematic variable 
and the ensuing iterated integrals are given by

\begin{equation}
    I(f_{i_n},\hdots,f_{i_1};x) = \int_{0}^x \mathrm dx_n\, f_{i_n}(x_n) \hdots \int_{0}^{x_3} \mathrm dx_2\, f_{i_2}(x_2) \int_{0}^{x_2} \mathrm dx_1 \,f_{i_1}(x_1)\,,
\end{equation}
where we fixed the boundary point at $x_0 = 0$ and we assume that
shuffle regularization is used throughout to make sense of the
divergent differential forms.

As already hinted at in the previous section, in the case of the three-loop self-energy in QED, 
we find that the space of differential one-forms
is larger than just $\mathrm d\!\log$-forms, and differential one-forms related to 
the elliptic curve of the two-loop equal mass sunrise integral appear.
In order to find an $\epsilon$-factorized basis, we proceed in two equivalent ways: first, using
an ansatz as elucidated in refs.~\cite{Pogel:2022yat,Pogel:2022ken,Pogel:2022vat}, and second, employing the algorithm described in ref.~\cite{Gorges:2023zgv}.
We verified explicitly that both approaches generate the same $\epsilon$-factorized basis, up to a rotation by a constant matrix. For the $\mathrm d\!\log$-type integrals, we use a standard approach~\cite{ArkaniHamed:2010gh,Henn:2014qga,Henn:2020lye} to derive a canonical $\epsilon$-factorized basis, including a leading singularity analysis in Baikov representation~\cite{Baikov:1996iu,Baikov:1996rk,Harley:2017qut}. We write our basis compactly in appendix~\ref{app:can} and an explicit version expressed in terms of the 
basis integrals in eq.~\eqref{eq:misI} can be found in the ancillary file to this manuscript.

We write our $\epsilon$-factorized equations as in eq.~\eqref{eq:diffeps}, and we find that we need 
a total of $16$ differential forms $\omega_i = f_i(x) \mathrm dx$. $9$ of these differential forms are $\mathrm d\!\log$ forms, and the remaining $7$
involve a solution of the Picard-Fuchs equation~\eqref{eq:pfell}. By using the notation in eq.~\eqref{eq:GMf}, the functions which determine
the differential forms up to three loops can be written as
\begin{align}
f_i \in \Bigg\{ &\frac1{x+3},   \frac1{x+1}, \frac1x, \frac1{x-1}, \frac1{x-2}, \frac1{x-9}, \nonumber 
\\
               &\frac1{\sqrt{(3+x)(1-x)}}, \frac1{\sqrt{(1-x)(9-x)}}, \frac1{\sqrt{(1-x)(9-x)}}\frac1x \Bigg\} 
               \quad \mbox{for}\;  i =1,\hdots,9 \,,     \label{eq:lettdlog} \\
f_i \in \Bigg\{ 
    & \frac1{x (x-1)(x-9) \varpi_0(x)^2}, \varpi_0(x), \frac{\varpi_0(x)}{x-1}, 
    \frac{(x-3)\varpi_0(x)}{\sqrt{(1-x)(9-x)}},
    \frac{(x+3)^4\varpi_0(x)^2}{x(x-1)(x-9)}, \nonumber \\
    &\frac{(x+3) (x-1) \varpi_0(x)^2}{x(x-9)}, 
    \frac{\varpi_0(x)^2}{(x-1)(x-9)} \Bigg\}\quad \mbox{for}\; i =10,\hdots,16 \,,
    \label{eq:lettell}
\end{align}
where $\varpi_0(x)=\varpi_0^{[x_0]}(x)$ is a solution of the differential equation~(\ref{eq:pfell}) and holomorphic at the regular singular point $x_0$.
Explicitly, they are given in~\cref{eq:defper0,eq:defper1,eq:defper9,eq:defperinf}.
To simplify the notation, 
we suppress here and in what follows the dependence of the
solution $\varpi_0^{[x_0]}(x)$ on the singular point $x_0$.
The formulas in~\cref{eq:lettell} 
become explicit only once considered locally, close
to a given singular point. For each singular point, we use the
definitions of the holomorphic solutions given in~\cref{eq:defper0,eq:defper1,eq:defper9,eq:defperinf}.
We provide the corresponding numerical matrices $G_i$ in the ancillary files attached to the arXiv submission of this paper.

The differential one-forms $\omega_i$ with $1\le i\le9$ are $\mathrm d\!\log$-forms.
This is entirely obvious for $i\le6$, while the remaining three ones can be written as
\begin{equation}
    \begin{aligned}
        \omega_7 & = \frac{1}{2i} \mathrm d\!\log\left( \frac{1+x-ir_2}{1+x+ir_2} \right), \\
        \omega_8 & = \frac{1}{2} \mathrm d\!\log\left(\frac{\left(x-3+r_1\right)\left(x+3+r_1\right)}{\left(x-3-r_1\right)\left(x+3-r_1\right)}\right), \\
        \omega_9 & = \frac{1}{6} \mathrm d\!\log\left(\frac{\left(x-3+r_1\right)\left(x+3-r_1\right)}{\left(x-3-r_1\right)\left(x+3+r_1\right)}\right),
    \end{aligned}
\end{equation}
where we introduced
\begin{eqnarray}
    r_1 \; = \; \sqrt{\left(1-x\right)\left(9-x\right)},
    & &
    r_2 \; = \; \sqrt{\left(1-x\right)\left(3+x\right)}.
\end{eqnarray}
If we define $\eta_j(\tau)$ by
\begin{equation}
    \begin{aligned}\label{eq:eta_f_def}
          2 \pi i \; \eta_j\left(\tau\right) \mathrm d\tau & = & f_j(x) \; \mathrm dx,
    \end{aligned}
\end{equation}
with $\tau$ defined by eq.~(\ref{eq:tau}),
then $\eta_3(\tau), \eta_4(\tau), \eta_6(\tau), \eta_{10}(\tau) \eta_{11}(\tau), \eta_{12}(\tau), \eta_{14}(\tau), \eta_{15}(\tau)$ and $\eta_{16}(\tau)$ are modular forms of $\Gamma_1(6)$. We will comment more on the connection to modular forms in section~\ref{sec:modular}.

At this point, a comment on the independence of the differential forms is in order.
It is easy to see that  the $\mathrm d\!\log$-forms and the modular ones described above are independent, 
as they provide a basis for the corresponding cohomology groups. 
One could study the linear independence of these meromorphic modular forms using the methods of ref.~\cite{Broedel:2021zij}. 
It is easy to verify that no relation of the following type exists among the differential forms
\begin{align}
    \sum_{i=1}^{16} \alpha_i \omega_i = \mathrm d\left( a(x) (\varpi_0(x))^n \right) \quad \mbox{with} \quad \alpha_i \in \mathbb{C} \quad \mbox{and} \quad n \in \mathbb{Z}\,,\label{eq:formindep}
\end{align}
where $a(x)$ is an algebraic function of $x$.
As $\varpi_0$ and $\partial_x\varpi_0$ are transcendental and independent, the absence of $\partial_x \varpi_0$ on the left-hand side implies $n=0$. We then only have to deal with the dlog-forms.
Clearly, a logarithm cannot be expressed as an algebraic function, hence it follows that
$\alpha_i=0$ for all $i$.

A special case of eq.~(\ref{eq:formindep}) is the statement that there are no relations of the form
\begin{align}
    \sum_{i=1}^{16} \alpha_i \omega_i = 0 \,, 
    \quad \mbox{with} \quad \alpha_i \in \mathbb{C}\,. \label{eq:formindep2}
\end{align}
This is the statement that the $\omega_i$'s form a basis of a vector space.
Equation~(\ref{eq:formindep2}) as well as \eqref{eq:formindep} can alternatively be proved directly by comparing the local series expansion
representations. Up to 200 orders in the expansion of the differential forms $\omega_i$ and Frobenius solutions $\varpi_0$ and $\partial_x \varpi_0$, we have not found any relation of these types.
We notice that these relations do not exclude more subtle relations, where linear combinations of the differential forms 
could be expressed as a total differential of an arbitrarily complicated function. 
Nevertheless, we believe that the conditions in eq.~\eqref{eq:formindep} should be enough to guarantee that no combination of the iterated
integrals originating from these forms
evaluates trivially to zero or can be reduced to a lower length iterated integral over the same set of forms.

We stress that a subset of the differential forms that enter the differential equations at three loops is enough to fully describe all master integrals
relevant to compute the
electron propagator to one and two loops in QED. In particular, at one loop only a subset of the
$\mathrm d\!\log$-forms appears
\begin{equation}
    f_i^{(1l)} \in \Bigg\{ \frac1x, \frac1{x-1} \Bigg\}\,. \label{eq:lett1l}
\end{equation}
At two loops, instead, we already find elliptic differential forms. In particular, 
the full set of two-loop forms required to evaluate all two-loop master integrals to all 
orders in $\epsilon$ reads
\begin{equation}
    f_i^{(2l)} \in \Bigg\{  \frac1x, \frac1{x-1}, \frac1{x-9}, \frac1{x(x-1)(x-9)\varpi_0(x)^2}, \varpi_0(x), \frac{\varpi_0(x)}{x-1},\frac{(x+3)^4\varpi_0(x)^2}{x(x-1)(x-9)} \Bigg\}\,. \label{eq:lett2l}
\end{equation}
Note that, while these are all forms appearing in the differential equation matrix, this does not mean that
they all contribute to the self-energy at a given loop order, especially if we limit
ourselves to consider the self-energy only up to $\mathcal{O}(\epsilon^0)$. In fact, this is relevant
at three loops, where some forms in eqs.~\eqref{eq:lettdlog} and \eqref{eq:lettell} turn out to only contribute
at higher orders in $\epsilon$. We will elaborate more on this point in the next sections.

\section{Solution in terms of iterated integrals}
\label{sec:iterint}

With the differential forms provided in eqs.~\eqref{eq:lettdlog} and ~\eqref{eq:lettell},
it is straightforward to write formal solutions for each master integral order-by-order in
$\epsilon$ in terms of (Chen) iterated integrals, up to suitable boundary conditions. 
As it can be inferred from the differential forms, the differential equations have singularities 
at $x = \{ -3,-1,0,1,2,9,\infty \}$ and, in general, it is necessary to obtain different
local analytic results in each region bounded by two of the singular points.
A formal solution in terms of iterated integrals 
can be written in each of those regions, and the corresponding boundary constants
can be fixed using different methods, which include the use of regularity conditions and 
direct calculation of some master integrals for special values of the kinematics.
Typically, one aims at determining all boundary conditions at one selected point (for example, $x=0$),
and write a solution which is valid close to that point. This solution can then be analytically continued to the 
whole kinematic space, by transporting the boundary conditions to any other singular (or even non-singular) point. 
Alternatively, one can recompute different boundary values in each region (numerically or analytically), and write a new
solution valid in each of them from scratch. Both approaches have their pros and cons. We
 describe our procedure for the analytic continuation in detail in section~\ref{sec:global}. 

Here, we first focus on determining the complete set of solutions for all master integrals 
in terms of iterated integrals and boundary conditions in one region. It is natural to consider four special 
points as
boundary values, i.e. $p^2=\{0, m^2, 9m^2,\infty\}$, or equivalently $x=\{0,1,9, \infty\}$. In fact, these are all
singular points of the Picard-Fuchs equation~\eqref{eq:pfell}, and so they correspond to points where the elliptic curve in eq.~\eqref{eq:curve} degenerates. We therefore expect 
that the boundary constants at those points should be polylogarithmic.
Out of the three points above, $p^2 = m^2$ (i.e., $x=1$) plays a special role.
In this limit the electron goes on-shell, and we expect its self-energy
to be expressible in terms of the same set of transcendental numbers relevant for the calculation of the
electron form factor at zero momentum in QED, i.e., the $g$-2 of the electron and to the so-called Dirac slope~\cite{Laporta:1996mq,Melnikov:1999xp}. These are known to be expressible up to three loops in terms of multiple zeta values and classical polylogarithms evaluated at $\tfrac{1}{2}$, i.e., only transcendental numbers from the following set appear:
\begin{equation}
    \left\{  \pi,\, \log(2),\, \zeta(3),\, \text{Li}_4(\tfrac{1}{2}),\, \zeta(5) \right\} \, .
\end{equation}
Notice that in our master integrals also the transcendental number $\text{Li}_5(\tfrac{1}{2})$ appears. This number drops out in the self-energy up to $\mathcal O(\epsilon^0)$, as it does so for the computation of the $g$-2 of the electron. 

While we could recover most boundary values close to $x=1$ from the analytic results for the master integrals
provided in refs.~\cite{Laporta:1996mq,Melnikov:1999xp}, we decided to follow a different
approach that can be more easily generalized to other boundary points as well. 
In order to obtain the limiting value of the master integrals close to a regular singular point $x=x_0$,
we use \texttt{AMFlow}~\cite{Liu:2017jxz,Liu:2022chg} to obtain a high precision evaluation
close to that point. For our problem, we typically choose $x = x_0 \pm \delta$ with $\delta \sim \mathcal{O}(10^{-2})$. 
We can then use the differential equations to transport this boundary value back to exactly $x=x_0$.
Let us assume for definiteness that we perform the matching to the right of the singular point.
Then we achieve this by computing, order by order in $\epsilon$, 
a generalized series expansion solution for the master integrals close to $x=x_0^{+}$ with 
an arbitrary number of terms and matching this to the numerical result obtained at $x = x_0+\delta$.
In formulas, if we expand the master integrals in $\epsilon$ as
\begin{equation}
    \vec{J}(x) = \sum_{n=0}^\infty \epsilon^n\, \vec{J}^n(x)\,,
\end{equation}
we can write
\begin{equation}
    \vec{J}^{(n)}(x) = \sum_{a\in I_a} \sum_{b\in I_b} \vec{C}_{a,b}^{(n)}\,  (x-x_0)^{a} \log^b{(x-x_0)}\,, \label{eq:expJ}
\end{equation}
where $a$ and $b$ run over the sets of indices $I_a$ and $I_b$, respectively. 
Note that $I_b$ is fixed at every order in $\epsilon$ by the differential equations, 
while $I_a$, which in general also includes half-integers, 
can be extended arbitrarily to increase the precision of the series expansion.
In order to determine high-precision numerical values for the constants $\vec{C}_{a,b}^{(n)}$, we
match them numerically by evaluating the series expansion at $x=x_0+\delta$ and
imposing that it agrees with the result obtained with \texttt{AMFlow}.
By repeating the same procedure with an increased number of terms in the series expansion, we can check that we have correctly
transported back to $x=x_0^{+}$ all digits evaluated at $x=x_0+\delta$.

We mention that, if required, a similar procedure can be adapted to obtain a fully $\epsilon$-resummed result for the 
master integrals close to any regular-singular point, by which 
we mean that their singular behavior close to the
singular point was \emph{resummed} in the dimensional regulator $\epsilon$ into the following form
\begin{equation}
    \vec{J}(x) = \sum_{i,j,k} \vec{D}_{i,j,k}(\epsilon) (x-x_0)^{i+j \epsilon} \log^k{(x-x_0)} \,.
\label{eq:expJres}
\end{equation}
Again the ranges for the indices $j,k$ are finite and completely fixed by the differential 
equations\footnote{Note that in general $j$ does not need
to be positive.} while $i$, which in general ranges over half-integers, can be increased to improve 
the precision of the expansion. We stress that the existence of logarithms
which are not generated by expanding terms of the form $(x-x_0)^{i+j \epsilon}$ in $\epsilon$
is a general feature of asymptotic expansions in dimensional regularization. Note that 
the $\vec{D}_{i,j,k}(\epsilon)$ are now 
functions of $\epsilon$ (and, for example, could include poles in the dimensional regulator $\eps$, even if the
original master integrals are finite). 

We will mainly content ourselves with obtaining results as in eq.~\eqref{eq:expJ}, which are enough
to fully determine the master integrals close to any singular point.
The point $p^2 = m^2$ ($x=1$) is special. There, we also obtained a fully $\epsilon$-resummed result as in eq.~\eqref{eq:expJres},
which is especially useful to recompute the QED mass and wave function renormalization constants in the on-shell scheme.
The wave function renormalization is in fact well known to be affected by infrared divergences in the limit $p^2 \to m^2$ ($x \to 1$), and a resummed result for the propagator allows us to compute it directly from our results, 
by simply discarding all divergent branches in the
expansion in eq.~\eqref{eq:expJres} with $x_0 = 1$. We will provide more details in the subsection~\ref{sec:ren}.

\subsection{General structure of the solution in terms of iterated integrals}
\label{sec:gen_struc}
Starting from our $\epsilon$-factorized basis of differential equations, supplemented
with analytic boundary values at $x=1$, it is straightforward to obtain iterated integral representations for all the master integrals.
For definiteness, we focus on the region $0<x<1$, which allows us to probe
two of the interesting limiting points $x\to 0^+$ and $x\to 1^-$. The final expressions are compact, 
but still too lengthy to be reported here.
We make them available as ancillary files attached to the arXiv submission of the manuscript.
In what follows, we discuss some of their properties in more detail, which become manifest order by order 
in $\epsilon$, thanks to the $\epsilon$-factorized basis that we used 
for the calculation of the master integrals.
Once more, this is because the differential forms that we identified up to three loops are independent, see eq.~\eqref{eq:formindep},
and can be defined
to have only single poles locally close to any given singular point. We therefore  expect all
iterated integrals to be linearly independent. We will focus in particular on
the order $\mathcal{O}(\epsilon^0)$ of the electron self-energy up to three loops.

\subsubsection*{Differential forms contributing to the finite remainder} 
First of all, it is  interesting to note that, at each loop order,
    many of the complicated letters in~\cref{eq:lettdlog,eq:lettell,eq:lett1l,eq:lett2l},
    do not contribute to the order $\mathcal{O}(\epsilon^0)$ of the bare electron propagator.
    In particular, at one loop it is easy to see that only one kernel appears in the iterated integrals
    \begin{equation}
        \frac1{x-1}\,.\label{eq:kern1lphys}
    \end{equation}
    Similarly, at two loops the following subset of kernels appears
\begin{equation}
   \left\{ \frac1x,\, \frac1{x-1},\, \frac1{\Delta \varpi_0(x)^2},\, \varpi_0(x),\, \frac{\varpi_0(x)}{x-1} \right\}\,,\label{eq:kern2lphys}
\end{equation}
    where $\Delta = x(x-1)(x-9)$ is the discriminant associated to the Picard-Fuchs equation~\eqref{eq:pfell}. 
     Finally, at three loops one logarithmic kernel and three elliptic kernels drop out, and we are left with a 
     subset of $12$ of the original $16$ kernels
\begin{equation}
\begin{aligned}
    & \left\{ \frac1x,\, \frac1{x-1},\, \frac1{x+1},\, \frac1{x-2},\, \frac1{x-9},\, \frac1{\sqrt{(3+x)(1-x)}},\, \frac1{\sqrt{(1-x)(9-x)}}, 
    \right.\\ 
    & \left. \frac1{\sqrt{(1-x)(9-x)}}\frac1x, \,
     \frac1{\Delta \varpi_0(x)^2},\, \varpi_0(x),\, \frac{\varpi_0(x)}{x-1},\, \frac{(x-3)\varpi_0(x)}{\sqrt{(1-x)(9-x)}} \right\}\, .
\end{aligned}
\end{equation}

These results deserve some discussion. 
It turns out that both at two and three loops, where iterated integrals of elliptic type make their
appearance, all kernels containing positive powers of $\varpi_0^2(x)$ drop from the finite part of the electron self-energy.
This can be understood as follows: 
Every elliptic sector contains two master integrals, which form the elliptic system.
The first master integral is normalized by $\varpi_0$, the second master integral
is the derivative of the first one plus a remainder term.
From the structure of the differential equation it is clear that kernels
containing positive powers of $\varpi_0^2(x)$ occur for the first time 
in the second master integral at the order in $\eps$, which is one higher than the first
non-vanishing order in $\eps$ of the first master integral.
As the second master integral involves the derivative of the first one,
the second master integral has an integrand with double poles.

\subsubsection*{Differential forms contributing to the poles} 
From our analytic representation in terms of iterated integrals, 
it is also straightforward to see that at each loop order
an even smaller subset of iterated integrals contribute to the poles of the corresponding 
self-energy coefficients $\Sigma_N^{(\ell)}$.
In particular, as one can easily see from the analytic results provided in the ancillary files, 
the poles at two loops only involve the one-loop kernels in eq.~\eqref{eq:kern1lphys},
while the poles of the three-loop self-energy only involve the two-loop kernels in eq.~\eqref{eq:kern2lphys}. 
This is expected, as the expressions for the poles are related to lower loop results. 
The fact that our representation makes this manifest, 
also renders UV renormalization straightforward in practice, directly at the level of iterated integrals.

As exemplification, we can showcase the poles of the three-loop self-energy coefficients.
They are simple enough that they can be written down in few lines.
We expand the self-energy coefficients explicitly as 
\begin{equation}
\begin{aligned}
    \Sigma_N^{(3)} =   \frac{\Sigma_{N,-3}^{(3)}}{\epsilon^3} + \frac{\Sigma_{N,-2}^{(3)}}{\epsilon^2}+ \frac{\Sigma_{N,-1}^{(3)}}{\epsilon^1} +\mathcal O(\epsilon ^0)   
\end{aligned}
\end{equation}
and put
$\xi=0$ for definiteness. The poles can then be expressed as
\begin{align}
    \Sigma_{V,-3}^{(3)} &= 0 \, , \nonumber \\
    \Sigma_{V,-2}^{(3)} &= \frac7{144} \, , \nonumber\\
    \Sigma_{V,-1}^{(3)} &=  R_1 + R_2I(f_4;x) + \frac{R_3}{\varpi_0}I(f_{11};x) + R_4I(f_4,f_4;x) 
    + R_5 \Big[ 6I(f_3,f_4;x)-\pi^2 \Big] \nonumber \\
    & + \left( R_6\varpi_0+R_7\varpi_0' \right) \left[ 9I(f_{10},f_{11};x)
    -2\sqrt3 \text{Cl}_2(\tfrac{\pi}{3}) \right] + R_8 \Big[ 144 I(f_{12},f_{10},f_{11};x)  \nonumber \\
    &  +36I(f_3,f_4,f_4;x)-18I(f_{11},f_{10},f_{11};x)-18I(f_4,f_3,f_4;x) +3\pi^2I(f_4;x) \nonumber \\
    &  +4\sqrt3 \text{Cl}_2(\tfrac{\pi}{3}) \big( I(f_{11};x) - 8 I(f_{12};x)\big) \Big] \, , \nonumber\\
    \Sigma_{S,-3}^{(3)} &= -\frac{5}{1152} \, , \nonumber\\
    \Sigma_{S,-2}^{(3)} &= -\frac{673+56 x}{864 (1-x)}     -\frac{437-5 x}{384 x}I(f_4;x) \, , \nonumber\\
    \Sigma_{S,-1}^{(3)} &= \tilde R_1 + \tilde R_2 \Big[ 7\pi^2-144\zeta(3) \Big] 
    + \tilde R_3 I(f_4;x)  + \frac{\tilde R_4}{\varpi_0} I(f_{11};x) + \tilde R_5 I(f_4,f_4;x) \, , \nonumber\\
    & + \tilde R_6 \Big[ 7I(f_3,f_4;x)-24\zeta(3) \Big]
    + \left( \tilde R_7\varpi_0+\tilde R_8\varpi_0'\right) \Big[ 9I(f_{10},f_{11};x)
    -2\sqrt3 \text{Cl}_2(\tfrac{\pi}{3}) \Big] \nonumber \\
    & + \tilde R_9 \Big[ 1008 I(f_{12},f_{10},f_{11};x)+252I(f_3,f_4,f_4;x)-126I(f_{11},f_{10},f_{11};x) \nonumber\\
    & -126I(f_4,f_3,f_4;x)+28\sqrt3 \, \text{Cl}_2(\tfrac{\pi}{3}) \big( I(f_{11};x) -8  I(f_{12};x) \big) \nonumber\\
    & +21\pi^2I(f_4;x)-2484\zeta(3)\Big]\, ,
\end{align}
where $\varpi_0$ is the holomorphic solution given in~\cref{eq:defper0,eq:defper1,eq:defper9,eq:defperinf},
$\varpi_0'$ its derivatives and $R_i$, $\tilde R_i$ are simple
rational functions.
They read explicitly
\begin{align}
    R_1 &= \frac{6648+20855 x+9770 x^2+4199 x^3}{17280 (-1+x)^2 x}\,,   \quad 
    R_2 = \frac{-2623+810 x+84 x^2-26 x^3}{1440 x^2} \, , \nonumber \\
    R_3 &= \frac{1071-2571 x+344 x^2-438 x^3-147 x^4+13 x^5}{720 (x-1)^3 x^2}\,,  \quad 
    R_4 = \frac{-69-36 x+20 x^2+4 x^3}{48 x^3} \, , \nonumber \\
    R_5 &= \frac{(2+x) (7-4 x)}{144 (x-1) x}\,,   \quad 
    R_6 = \frac{(9-x) \left(777-896 x+516 x^2+48 x^3-13 x^4\right)}{6480 (x-1)^2 x} \, , \nonumber \\
    R_7 &= -\frac{(9-x) \left(1071-2571 x+344 x^2-438 x^3-147 x^4+13 x^5\right)}{6480 (x-1)^2 x}\,, \quad 
    R_8 = \frac{7}{216 x^2}  \, .
\end{align}
and
\begin{align}
    \tilde R_1 &= \frac{1485-15418 x-7711 x^2+908 x^3}{5184 (x-1)^2 x}\,, \quad 
    \tilde R_2 = -\frac{37+35 x}{8064 (x-1)} \, , \nonumber \\
    \tilde R_3 &= \frac{-660+3185 x-5749 x^2+1280 x^3}{1152 (x-1) x^2}\,, \quad 
    \tilde R_4 = \frac{171-336 x+547 x^2-94 x^3}{72 (x-1)^3 x} \, ,\nonumber \\
    \tilde R_5 &= \frac{110-444 x+859 x^2-93 x^3}{192 x^3}\,, \quad 
    \tilde R_6 = \frac{437-368 x+75 x^2}{2688 (x-1) x} \, ,\nonumber \\
    \tilde R_7 &= \frac{(9-x) \left(39-113 x+38 x^2\right)}{324 (x-1)^2}\,, \quad
    \tilde R_8 = \frac{(9-x) \left(-171+336 x-547 x^2+94 x^3\right)}{648 (x-1)^2} \, ,\nonumber \\
    \tilde R_9 &= \frac{19+x}{12096 x} \, .
\end{align}
Note that to obtain these expressions we have fixed the boundary values for the iterated integrals at $x=0$, 
which explains the appearance of the Clausen function $\text{Cl}_2\!\left(\tfrac{\pi}{3}\right)$ in our formulas, with $\text{Cl}_2\!\left(x\right) = \textrm{Im}\!\left(\textrm{Li}_2\!\left(e^{ix}\right)\right)$. 
It turns out that the boundary values are multiple polylogarithms up to weight 4 evaluated at sixth root of unity. 
A basis for this class of transcendental numbers has been derived up to weight 3 in ref.~\cite{Huber:2009se} and weight 6 in ref.~\cite{Henn:2015sem}. 
Here we use a different basis, which is chosen in such a way that as many transcendental numbers as possible are expressible in terms of classical polylogarithms. 
This becomes possible using motivic techniques, in particular the decomposition into primitives of ref.~\cite{brownperiods}, which generalizes the well-known $f$-alphabet for (motivic) multiple zeta values~\cite{brownMTZ,Brown:2011ik} to larger classes of (motivic) periods. Using this basis, we find that the full set of transcendental numbers (and products thereof) sufficient 
to fix all necessary boundary conditions at $x=0$ up to the order relevant to us is given by the set
\begin{equation}\label{eq:constants_at_3_loops}
    \{  \pi,\,\log(2),\,\log(3),\, \sqrt{3}\text{Cl}_2\!\left(\tfrac{\pi}{3}\right),\,\zeta(3),\, \text{Im}(\text{Li}_3(\tfrac{i}{\sqrt{3}})),\,\text{Li}_4\!\left(-\tfrac{1}{3}\right), \,\text{Li}_4\!\left(\tfrac{1}{2}\right), \, \text{Li}_4\!\left(\tfrac{1}{3}\right)   \} \, .
\end{equation}

After having discussed the general analytic properties of the self-energy
through its representation in terms of general iterated integrals, in the following sections, 
we will consider its limiting behavior close to various singular points, focusing  on $p^2 \to m^2$, 
which is of special physical significance.


\subsection{Local solution close to $p^2 = m^2$}
\label{sec:local1}

The behavior of the self-energy for $p^2 \to m^2$ is
physically relevant, as it is central to determine the mass and wave function 
renormalization constants in the on-shell scheme.
As it is well known, $p^2$ acts in this case as an infrared regulator,
and taking the on-shell limit produces new infrared divergences,
which render the extraction of the renormalization constants from
the full result non-trivial. 
As discussed above, see eq.~\eqref{eq:expJres}, from our $\epsilon$-factorized differential equations, 
it is  easy to obtain 
a fully $\epsilon$-resummed expression for the self-energy coefficients
close to $p^2 = m^2$. This makes the branch structure of the self-energy
manifest and also allows us to take the on-shell limit at once, simply by discarding
the divergent branches.

Following the definition in eq.~\eqref{eq:defSigmaVS}, we expect that the
behaviour of the scalar coefficients of the self energy close to $p^2 = m^2$, i.e. $x=1$,
can be written in general as
\begin{equation}
\Sigma_{N,\text{res}}^{(\ell)}(p,m) 
=
\sum_{i,j,k} S_{i,j,k}^{N,(\ell)}(\epsilon) (1-x)^{i+j \epsilon} \log^k{(1-x)}\,,
\end{equation}
with $N=V,S$. We assume that $0<x<1$ for definiteness.
We find that
the coefficients $S_{i,j,k}^{N,(\ell)}(\epsilon)$
are expressible in terms of multiple zeta
values and classical polylogarithms evaluated at $\tfrac{1}{2}$ up to transcendental weight $5$.
We provide the full results for the 
$S_{i,j,k}^{N,(\ell)}(\epsilon)$ up to three loops in the 
ancillary files, keeping full dependence on the gauge parameter $\xi$. 
Here, we are only interested in discussing  the general features of the solution at each loop order,
and to do that we showcase the first orders of the expansion.

At one loop, one finds two different asymptotic behaviors in the limit $x \to 1$
\begin{equation}
    \Sigma_{N,\text{res}}^{(1)} = 
\sum_{i=0}^\infty A_i^{N}(\epsilon) (1-x)^{i} 
+\sum_{i=1}^{\infty} B_i^{N}(\epsilon) (1-x)^{i-2 \epsilon}
\,, \label{eq:res1L}
\end{equation}
where we notice that the non-trivial branch 
$(1-x)^{-2 \epsilon}$ is suppressed by  a power of $(1-x)$.
Keeping only the first non-zero orders in $(1-x)$, and expanding the corresponding
functions in $\epsilon$, we find that the full 
 self-energy can be written as 
\begin{equation}
    \hat{\Sigma}_{\text{res}}^{(1)} = 
    \xi  \left[\frac{1}{4 \epsilon}+\frac{1}{2} + \mathcal{O}(\epsilon)\right] \slashed{p} 
    +\left[ \xi  \left(-\frac{1}{4 \epsilon
   }-\frac{1}{2}\right)-\frac{3}{4 \epsilon }-1 + \mathcal{O}(\epsilon)\right] m + \mathcal{O}(1-x)
\end{equation}

At two loops, we find instead three non-trivial branches
\begin{equation}
    \Sigma_{N,\text{res}}^{(2)} = 
\sum_{i=0}^\infty C_i^{N}(\epsilon) (1-x)^{i} 
+\sum_{i=0}^\infty D_i^{N}(\epsilon) (1-x)^{i-2 \epsilon}
+\sum_{i=1}^\infty E_i^{N}(\epsilon) (1-x)^{i-4 \epsilon}
\,.\label{eq:res2L}
\end{equation}
Also at two loops we notice that the highest branch  
$(1-x)^{-4 \eps}$ is suppressed by an extra power of $(1-x)$.
Keeping only the first non-trivial orders in $(1-x)$, and putting $\xi=0$
for simplicity, we can write 
\begin{equation}
\begin{aligned}
    \Sigma_{V,\text{res}}^{(2)} &=
    \left[ -\frac{7}{64 \epsilon }+\frac{91}{128}-\frac{\pi ^2}{12}+\frac{1}{4} \pi ^2 \log (2)-\frac{3 \zeta (3)}{8} + \mathcal O(\eps) \right] + \mathcal O(1-x) \\
    \Sigma_{S,\text{res}}^{(2)} &=
    \left[ -\frac{23}{32 \epsilon ^2}-\frac{5}{8 \epsilon }-2-\frac{19 \pi ^2}{48}+\frac{1}{4} \pi ^2 \log (2)-\frac{3 \zeta
   (3)}{8} + \mathcal O(\eps) \right] \\
   &+(1-x)^{-2 \eps} \left(
   \frac{9}{16 \epsilon ^2}+\frac{3}{8\epsilon}
 +1
   -\frac{15 \pi ^2}{16}+ \mathcal O(\eps)\right) + \mathcal O(1-x) \,,
\end{aligned}    
\end{equation}
where we recall that we assumed $0<x<1$ throughout.

Expanding in $\epsilon$ and retaining only the logarithmically
enhanced terms and the constant part in $(1-x)$,
the expressions simplify further and the bare 
self-energy at two loops  can be written in compact form as
\begin{align}
    \Sigma_{V}^{(2)} =  &-\frac{7}{64 \epsilon}+\frac{91}{128}
    -\frac{\pi ^2}{12} +\frac{1}{4} \pi ^2 \log(2) -\frac{3 \zeta (3)}{8}
    + \mathcal{O}(\epsilon)\,, \\
    \Sigma_{S}^{(2)} = & \phantom{+} \frac{9}{8} \log ^2(1-x)  
  +\left(-\frac{9}{8 \epsilon }-\frac{3}{4}\right)\log (1-x) \nonumber \\ &
   -\frac{5}{32 \epsilon ^2}-\frac{1}{4 \epsilon} -1-\frac{5 \pi ^2}{24}
   +\frac{1}{4} \pi ^2 \log (2) -\frac{3 \zeta(3)}{8} + \mathcal{O}(\epsilon)\,.
\end{align}
We stress that, in order to match the transcendental weight of the three-loop results, 
we derived results up to transcendental weight $5$ also for the one- and two-loop coefficients,
which are provided in the ancillary files that accompany the arXiv submission of this paper.

Finally, at three loops the branch structure is richer and the resummed self-energy
can be expressed as
\begin{align}
    \Sigma_{N,\text{res}}^{(3)} &= 
\sum_{i=0}^\infty F_i^{N}(\epsilon) (1-x)^{i} 
+\sum_{i=-1}^\infty G_i^{N}(\epsilon) (1-x)^{i-2 \epsilon}  \nonumber \\
&
+\sum_{i=0}^\infty H_i^{N}(\epsilon) (1-x)^{i-4 \epsilon}
+\sum_{i=1}^\infty J_i^{N}(\epsilon) (1-x)^{i-6 \epsilon}
\,, \label{eq:res3L}
\end{align}
where we see once more that the highest branch $(1-x)^{-6 \eps}$ is  
suppressed by a power of $(1-x)$. Moreover, we notice that the branch $(1-x)^{-2 \eps}$
starts with a single pole as $x\to 1$. 
We will elaborate more on this in the following.
As exemplification, keeping just the first orders in $\epsilon$, we can write
\begin{equation}
\begin{aligned}
    \Sigma_{V,\text{res}}^{(3)} &= 
    \left[ -\frac{27}{128 \epsilon ^3}-\frac{673}{1152 \epsilon ^2} 
    + \mathcal O\left(\frac1\epsilon\right) \right] %
    + (1-x)^{-2 \eps}\left[
    \frac{27}{64 \epsilon ^3}+\frac{27}{32 \epsilon ^2}
    + \mathcal O\left(\frac1\epsilon\right)\right] \\
    &\quad+ (1-x)^{-4 \eps}\left[
    -\frac{27}{128 \epsilon ^3}- \frac{27}{128 \epsilon ^2}
    + \mathcal O\left(\frac1\epsilon\right)\right] + \mathcal O(1-x) \, , \\
    \Sigma_{S,\text{res}}^{(3)} &= \left[ 
    -\frac{653}{1152 \epsilon ^3}+\frac{1447}{6912 \epsilon ^2} + \mathcal O\left(\frac1\epsilon\right) \right] %
    + (1-x)^{-2 \eps}\left[
     \frac1{x-1} \left( \frac{27}{32 \epsilon ^2}+ \mathcal O\left(\frac1\epsilon\right) \right) \right.\\ 
    &\quad \left.+ \frac{9}{16 \epsilon ^3} -\frac{91}{256\epsilon ^2}
    + \mathcal O\left(\frac1\epsilon\right)
    \right]+ (1-x)^{-4 \eps}\left[
    \frac{27}{128 \epsilon ^2}+ \mathcal O\left(\frac1\epsilon\right)
    \right]+ \mathcal O(1-x) \,,
\end{aligned}    
\end{equation}
while the complete results to order $\mathcal{O}(\epsilon^0)$ 
for the various branches can be obtained in computer
readable format from the arXiv submission of this paper.
If we  expand all branches in $\epsilon$ to $\mathcal{O}(\epsilon^0)$, 
and retain only the non-suppressed terms in $(1-x)$, 
we find the rather compact expressions
\allowdisplaybreaks
\begin{align}
\Sigma_{V}^{(3)} &=  
    \frac{27}{16} \log^3(1-x)
    -\frac{27 \log^2(1-x)}{32 \epsilon }  +\left(-\frac{27}{32 \epsilon }
   -\frac{27}{8} +\frac{9\pi ^2}{32}
   \right) \log (1-x)  \nonumber \\
    & 
   +\frac{7}{144 \epsilon ^2} 
   + \frac{1}{\epsilon } \left(
   \frac{5179}{3456}-\frac{139 \pi ^2}{576}
   +\frac{7}{12}\pi ^2 \log (2)-\frac{7 \zeta(3)}{8} \right) 
   \nonumber \\
    & + \frac{5815}{20736} -\frac{6827 \pi^2}{12960} 
     -\frac{5149\zeta (3)}{576}
    +\frac{373}{96}\pi ^2 \log (2) 
    +\frac{433 \pi^4}{3456}-\frac{83 \log^4(2)}{144}
    \nonumber \\
    &
   -\frac{11}{18} \pi ^2 \log ^2(2) -\frac{83}{6}  \text{Li}_4\left(\tfrac{1}{2}\right)
   +\frac{5 \zeta (5)}{16} + \mathcal{O}(\epsilon) \,, \label{eq:sigma3lVone}\\ 
\Sigma_{S}^{(3)} &=   
   \frac{1}{x-1}\left[
   \frac{27}{16} \log^2(1-x)
   +\left(-\frac{27}{16 \epsilon} -\frac{27}{8}\right) \log (1-x)
   +\frac{27}{32 \epsilon ^2}
   +\frac{27}{16 \epsilon}+\frac{9}{2}+\frac{9 \pi ^2}{32}\right] \nonumber \\
   &-\frac{3}{4}\log ^3(1-x)
   +\left(\frac{9}{8\epsilon }+\frac{125}{128}\right) \log ^2(1-x)
     \nonumber \\
   &  + \log (1-x)\left(
   -\frac{9}{8 \epsilon^2}-\frac{17}{128 \epsilon }-\frac{3433}{768}
   -\frac{53 \pi^2}{32}
   +\frac{3}{4} \pi ^2 \log(2) -\frac{9 \zeta (3)}{8}\right)  
     \nonumber \\
   & -\frac{5}{1152\epsilon ^3}+\frac{7}{108 \epsilon ^2}
   +\frac{1}{\epsilon }\left(\frac{8117}{5184}-\frac{95 \pi ^2}{288}+\frac{5}{24}\pi ^2 \log (2)
   -\frac{9 \zeta(3)}{16} \right) \nonumber \\
&    +\frac{162239}{31104} -\frac{768617 \pi^2}{103680}
     +\frac{3139}{288}\pi ^2 \log (2) -\frac{13237 \zeta (3)}{576}  
   +\frac{997 \pi^4}{5760}  \nonumber \\
   & 
   -\frac{29 \log^4(2)}{48}
   - \frac{7}{12} \pi ^2 \log ^2(2) 
   -\frac{29\text{Li}_4\left(\tfrac{1}{2}\right)}{2}
   -\frac{\pi^2 \zeta (3)}{16} +\frac{5 \zeta (5)}{16}+ \mathcal{O}(\epsilon)\,. \label{eq:sigma3lSone}
\end{align}

As hinted to before, the analytic result for $\Sigma_S^{(3)}$ 
in eq.~\eqref{eq:sigma3lSone} deserves some discussion, due to the presence
of a single pole in the limit $x \to 1$, in addition to the usual logarithmic singularities.
While it might appear surprising, this behavior is to be expected, 
and can be traced back to the class of three-loop
diagrams which contain two-loop \textsl{one-particle reducible} subgraphs. 
This phenomenon appears for the first time
at three loops. We expect that, upon renormalization in the on-shell scheme, this
pole should disappear in order to preserve the correct renormalization conditions, i.e.,
the position of the pole at the physical electron mass, with residue equal to unity.
We will discuss  renormalization explicitly in section~\ref{sec:ren}.

\subsection{Local solution close to $p^2 = 0$}
\label{sec:local0}

As another example of the explicit analytic results that we can obtain from our calculation, we study the limiting
behavior of the self-energy in the zero momentum limit, $x \to 0$.
This limit does not present any subtleties compared to the previous one. In particular,
 all master integrals are finite in the limit $x\to 0$, and we also expect the final
result for the self-energy to be finite in this limit.
Note that, despite this, it in not enough to compute only the zeroth order in the expansion
close to $x=0$, because there are poles in $1/x$, which appear in the coefficients of the 
decomposition of the amplitude in terms of our basis of master integrals.
Interestingly, it turns out that in this limit all master integrals reduce to combinations
of multiple polylogarithms evaluated at sixth roots of unity. As already discussed in section~\ref{sec:gen_struc}, using motivic techniques
we have determined a basis for this class of transcendental numbers up to weight 4 that consists as much as possible of classical polylogarithms. As a consequence, the transcendental numbers appearing in our result are independent. In particular, this means that, if there are cancellations between transcendental numbers of a certain type, they will be explicit in our basis. We indeed see that, even though individual master integrals involve transcendental constants which cannot be expressed in terms of classical polylogarithms, those all cancel in the contribution to the self-energy at three loop, and we only find transcendental constants that are special values of classical polylogarithms appearing through three loops up to finite terms, (cf.~eq.~\eqref{eq:constants_at_3_loops}). 
Specifically, assuming for definiteness 
$p^2 >0$ and $x>0$, we find (as for the previous limit, 
we provide results for the
gauge parameter $\xi = 0$ for simplicity, except for the one-loop case)

\begin{align}
     \Sigma_{V}^{(1)}|_{p^2 = 0} &=
    \xi \left( \frac{1}{4\epsilon} +\frac1 8     \right) +\mathcal{O}(\eps) \,,
      \\
    \Sigma_{S}^{(1)}|_{p^2 = 0}&=
    -\frac{3+\xi }{4\epsilon} -\frac{1+\xi }{4} +\mathcal{O}(\eps)   \,,
\end{align}
\begin{align}
     \Sigma_{V}^{(2)}|_{p^2 = 0}  &=
    -\frac{7}{64\epsilon} - \frac{211}{128}-\frac{\pi ^2}{16}
    +\frac{3\sqrt{3}}{2}\text{Cl}_2\!\left( \tfrac{\pi}{3} \right) +\mathcal{O}(\eps) \,,
      \\
     \Sigma_{S}^{(2)}|_{p^2 = 0}  &=
    -\frac{5}{32\epsilon^2}+\frac{19}{16\epsilon} +\frac{31}{8}-\frac{\pi ^2}{96}
    -\frac{9\sqrt{3}}{4}\text{Cl}_2\!\left( \tfrac{\pi}{3} \right)  +\mathcal{O}(\eps) \,,
\end{align}     

\begin{align}
\Sigma_{V}^{(3)}|_{p^2 = 0}  &= 
    \frac{7}{144\epsilon^2} +\left[\frac{5053}{3456}+\frac{\pi ^2}{24}
    -\sqrt{3}\text{Cl}_2\!\left( \tfrac{\pi}{3} \right) \right]\frac1\epsilon 
    +\frac{201881}{25920}+\frac{733 \zeta (3)}{24} 
     \nonumber \\
    & +\frac{5 \log ^4(2)}{3}
    -\frac{3 \log^4(3)}{16}-\pi ^2 \left[\frac{23}{360}
    +\frac{5 \log ^2(2)}{3}-\frac{3 \log ^2(3)}{8} +\frac{1459 \pi ^2}{8640}  \right] 
    \nonumber \\
   &    
   +\frac{15}{4}\text{Cl}_2\!\left( \tfrac{\pi}{3} \right)^2
   +40\text{Li}_4 \left( \tfrac{1}{2} \right)+\frac92\left[ \text{Li}_4 \left( -\tfrac{1}{3} \right) 
   - 2 \text{Li}_4\left( \tfrac{1}{3} \right)  \right]   \nonumber \\
   &  - \sqrt{3} \left[ \frac{1103  \pi ^3}{2160}
   + \frac{8197 }{480}\text{Cl}_2\!\left( \tfrac{\pi}{3} \right)  
   +\frac{23}{80}  \pi  \log ^2(3)
    -\frac{69}{5}\text{Im}\!\left( \text{Li}_3\!\left( \tfrac{i}{\sqrt3} \right) \right) \right]+\mathcal{O}(\eps)\,,
   \\
    \Sigma_{S}^{(3)}|_{p^2 = 0}  &=
   -\frac{5}{1152\epsilon^3}+\frac{1241}{3456\epsilon^2}
   - \left[\frac{3373}{1296}+\frac{\pi ^2}{1152}+\frac{\zeta (3)}{4} 
   +\frac{3 \sqrt{3}}{16}\text{Cl}_2\!\left( \tfrac{\pi}{3} \right)  \right]\frac1\epsilon  \nonumber \\ 
   &- \frac{4447957}{311040}  -\frac{7031 }{576}\zeta (3)   +\pi ^2 \left(\frac{3019}{17280}
   +\frac{7 \log^2(2)}{6}
   -\frac{3 \log ^2(3)}{16}    +\frac{151 \pi ^2}{1152} \right)
      \nonumber \\
   &  
   -\frac{21}{8} \text{Cl}_2\!\left( \tfrac{\pi}{3} \right)^2  
   -28\text{Li}_4(\tfrac{1}{2})-\frac{9}{4} \left[\text{Li}_4(-\tfrac{1}{3})
   -2\text{Li}_4(\tfrac{1}{3}) \right] -\frac{7 \log ^4(2)}{6}
   +\frac{3 \log ^4(3)}{32}     \nonumber \\
   & +\sqrt{3} \left[ \frac{691 \pi ^3}{960 } +\frac{22}{5}\text{Cl}_2\!\left( \tfrac{\pi}{3} \right)
    +\frac{129}{320}  \pi  \log ^2(3)
    - \frac{387  }{20}\text{Im}\!\left( \text{Li}_3\!\left( \tfrac{i}{\sqrt3} \right) \right)
   \right]+\mathcal{O}(\eps)\,
\end{align}
Let us make some comment about the appearance of $\sqrt{3}$ starting from two loops. At least up to three loops, the self-energy at $p^2=0$ can be expressed in terms of multiple polylogarithms evaluated at sixth roots of unity, i.e., multiple polylogarithms where all arguments are either 0 or $\rho^k$, with $0\le k<6$ and $\rho = e^{i\pi/3}=\tfrac{1}{2}+\tfrac{i\sqrt{3}}{2}$. Consequently, it is natural to separate the transcendental numbers into their real and imaginary parts. This defines a natural notion of parity on these numbers, under which the real parts are even, and the imaginary parts are odd. If we define $\sqrt{3} = 2\,\textrm{Im}(\rho)$ to be odd under this parity (and rational numbers are even), we see that the self-energy is even. The fact that the self-energy is parity-even can be seen as follows: The Feynman integrals that define the self-energy at a given loop order are integrals of rational functions, and thus independent of the choice of the sign of any square root that may appear in the answer. In particular, it cannot depend on the choice of sign for the square root of 3, and thus it is parity-even.

\subsection{Comments on the UV Renormalization}
\label{sec:ren}
As a cross-check of our result, we have used the $\eps$-resummed expressions
obtained in section~\ref{sec:local1} in order to recompute the mass and wave function
renormalization constants to three loops, first computed in ref.~\cite{Melnikov:2000zc}.
Following this reference, we renormalize the bare mass $m$ and electron wave function $\psi$
multiplicatively 
\begin{equation}
    m = Z_m \mr\,, \quad \psi = Z_2^{\frac{1}{2}}\,  \psir\,,
\end{equation}
where $\mr$ and $\psir$ are the renormalized quantities.
It is convenient to re-arrange the vector and scalar
form factors of the self-energy defined in eq.~\eqref{eq:defSigma} as follows
\begin{equation}
    \hat{\Sigma}(p,m) = \Sigma_V(p,m) \slashed{p}  + \Sigma_S (p,m)\, m  
    = \Sigma_1 (p,m)\, m  + \Sigma_2(p,m) (\slashed{p}-m) \,, 
\label{eq:defS1S2}
\end{equation}
where
\begin{equation}
    \Sigma_1(p,m) = \Sigma_S (p,m) + \Sigma_V (p,m) \,, \qquad \Sigma_2(p,m) = \Sigma_V(p,m)\,.
\end{equation}
After resumming one-particle irreducible graphs, 
the  renormalized two-point function can then be written as
\begin{align}
    \Pi_R(p) &= \frac{i Z_2}{\slashed{p} - Z_m \mr + \hat{\Sigma}(p,Z_m \mr )} \,.
     \label{eq:renProp}
\end{align}  
For ease of writing, but with a slight abuse of notation, we set from here on
$$\hat{\Sigma}(p,\mr) = \hat{\Sigma}(p,Z_m m_r)\,,$$
retaining just the explicit dependence on the renormalized mass.
Note that, since the only occurrences of the electron mass stem from the propagators of the
virtual electrons in the loops (external electrons are in general kept off-shell), it is
not necessary to use extra Feynman rules corresponding to the mass counterterms, instead it is
sufficient to renormalize the mass multiplicatively everywhere in the calculation of the self-energy.
Upon expanding in the coupling constant, this implies that we have to also consider mass derivatives of the 
self-energy.

The on-shell renormalization conditions imply that 
the propagator has a single pole at the on-shell mass $p^2 = \mr^2$
with residue normalized to $i$, i.e.,
\begin{equation}
    \lim_{p^2 \to \mr^2}\Pi_R(p) = \frac{i}{\slashed{p}-\mr}\,.
\end{equation}
To determine the renormalization constants in practice, 
we expand the self-energy close to $p^2 = \mr^2$. This allows us to write
\begin{align}
    \Pi_R(p) &=
    \frac{i Z_2}{
   (\slashed{p}-\mr) \left[ 1+ \Sigma_2|_{p^2=\mr^2} + 2\mr^2 \Sigma_1'|_{p^2=\mr^2} \right] +
    \mr\left[1-Z_m +\Sigma_1|_{p^2=\mr^2}\right] + ...} \label{eq:expprop}
\end{align}
where we defined 
$$\Sigma_j|_{p^2=\mr^2} =\Sigma_j(p,\mr)|_{p^2=\mr^2} \;\; \mbox{for} \;\; j=1,2\,, 
\qquad \Sigma_1'|_{p^2=\mr^2} = \frac{\partial}{\partial p^2}\Sigma_1(p,\mr)|_{p^2=\mr^2}\,,$$
and used the Taylor expansion for $\Sigma_1(p,\mr)$
\begin{equation}
    \Sigma_1(p,\mr) = \Sigma_1(p,\mr)|_{p^2=\mr^2} + 2\mr \frac{\partial}{\partial p^2}\Sigma_1(p,\mr)|_{p^2=\mr^2} (\slashed{p}-\mr) + ...\,.
\end{equation}
From eq.~\eqref{eq:expprop} one can easily read off the renormalization
conditions in the on-shell scheme  in terms of $\Sigma_1$ and $\Sigma_2$ as
\begin{equation}
\begin{split}
    Z_m &= 1 + \Sigma_1(p,\mr) \Big|_{p^2 = \mr^2} \,, \\ 
    \frac{1}{Z_2} &= 1 + 2 \mr^2 \frac{\partial}{\partial p^2} \Sigma_1(p,\mr)\Big|_{p^2 = \mr^2}
    + \Sigma_2(p,\mr)\Big|_{p^2 = \mr^2}\,. \label{eq:defZ}
\end{split}    
\end{equation}
Note that these equations involve the self-energy written in terms of the renormalized mass,
such that the dependence on the mass renormalization constant 
appears implicitly also in the right-hand side of these equations. 

Importantly, in order to properly take the on-shell limit in eq.~\eqref{eq:defZ}
 we must avoid spurious infrared divergences which
would otherwise spoil the result.
This can be achieved by using the $\eps$-resummed expressions provided in~\cref{eq:res1L,eq:res2L,eq:res3L} and
discarding all branches $\propto (1-x)^{j \epsilon}$ for $j \neq 0$. 
Of course, one must take special care in keeping enough orders in the expansion in $(1-x)$,
to properly account for the action of the derivative operator $\tfrac{\textrm{d}}{\textrm{d}x}$ in eq.~\eqref{eq:defZ}.

The renormalization constants
can be expanded in powers of the bare coupling constant as follows
\begin{align}
    &Z_m = 1 + \sum_{\ell=1}^\infty \left( \frac{\alpha}{\pi}C_r(\epsilon) \right)^\ell 
     Z_m^{(\ell)} \,, \nonumber \\
    &Z_2 = 1 + \sum_{\ell=1}^\infty \left( \frac{\alpha}{\pi}C_r(\epsilon) \right)^\ell  Z_2^{(\ell)}\,,
\end{align}
and using these expressions in eq.~\eqref{eq:defZ},
it is easy to recover the well known expressions for $Z_m^{(i)}$ and $Z_2^{(i)}$ for $i=1,2,3$,
first obtained in ref.~\cite{Melnikov:2000zc} 
and which we report in appendix~\ref{app:renconsts} for completeness.
We also defined, similarly to eq.~\eqref{eq:defCeps},
\begin{equation}
    C_r(\epsilon) = \Gamma(1+\epsilon) (4 \pi)^\epsilon (\mr^2)^{-\epsilon} \,.
    \label{eq:defCReps}
\end{equation}

After having reproduced all renormalization constants in the on-shell
scheme, we can use them to provide renormalized results for the self-energy up to three loops,
valid for any value of the momentum $p^2$.
Here we find it interesting to elaborate on the effect of mass 
renormalization on the self-energy (while we do not renormalize wave function and electric coupling).
Consistency with the renormalization conditions requires that the renormalized
self-energy must have the appropriate behaviour in the limit $x \to 1$. In particular,
the pole appearing in the three-loop scalar self-energy must cancel after
mass renormalization. We note that this check is non-trivial, since, as we argued explicitly, in deriving
the renormalization conditions, all divergent asymptotic behaviours $\propto (1-x)^{i \epsilon + j}$
have to be discarded.

We define the mass renormalized self-energy as follows
\begin{align}
    \hat{\Sigma}_{r}(p,\mr) = \hat{\Sigma}(p,Z_m \mr)\,, \label{eq:renSigma}
\end{align}
and write for convenience 
\begin{equation}
    Z_m = 1 + \delta m\,,
\end{equation}
where we can similarly expand
\begin{align}
    \delta m &= 1 + \sum_{\ell=1}^\infty \left( \frac{\alpha}{\pi} C_r(\epsilon)\right)^\ell  \delta m^{(\ell)}\,.
\end{align}
Expanding in the bare coupling $\alpha$ left- and right-hand sides of eq.~\eqref{eq:renSigma}
and collecting by powers of $\tfrac{\alpha}{\pi}$, 
we find  (we suppress the dependence on the arguments of the self-energy for compactness)
\begin{align}
\hat{\Sigma}^{(0)}_{r} &= \hat{\Sigma}^{(0)} \nonumber \\
\hat{\Sigma}^{(1)}_{r} &= \hat{\Sigma}^{(1)} -\mr\, \delta m^{(1)}\nonumber \\
\hat{\Sigma}^{(2)}_{r} &= \hat{\Sigma}^{(2)} -\mr\, \delta m^{(2 )}  
+ \mr\, \delta m^{(1)} \left(  \frac{\partial}{\partial m}\hat{\Sigma}^{(1)}  \right) \nonumber \\
\hat{\Sigma}^{(3)}_{r} &= \hat{\Sigma}^{(3)} -\mr\, \delta m^{(3)} 
+ \mr\, \delta m^{(2)} \left(  \frac{\partial}{\partial m}\hat{\Sigma}^{(1)}  \right)  \nonumber \\
& + \mr\, \delta m^{(1)} \left(  \frac{\partial}{\partial m}\hat{\Sigma}^{(2)}  \right) 
+ \frac{\mr^2}{2} \left(\delta m^{(1)} \right)^2 \frac{\partial^2}{\partial m^2}\hat{\Sigma}^{(1)}\,.
\label{eq:UVrenSigma}
\end{align}

While it is straightforward to perform renormalization 
for the self-energy written in terms of iterated integrals,
it is instructive to look at its behaviour close to $x = 1$. Here, we expect that
mass renormalization alone in the on-shell scheme should take care of removing all divergent
contributions, in particular the spurious pole appearing starting from three loops. This is necessary
to make sure that the full propagator has a pole at $p^2 = \mr^2$.
Explicitly, writing
\begin{equation}
    \Sigma_{r,V}(p,m) = \sum_{\ell=0}^\infty 
     \left( \frac{\alpha}{\pi}C_r(\epsilon) \right)^\ell    \Sigma_{r,V}^{(\ell)}(p,m)
\end{equation}
where $C_r(\epsilon)$ was defined in eq.~\eqref{eq:defCReps},
we find for the renormalized self-energy order by order in the loop expansion
that, as expected,
\beq
\lim_{p^2 \to m^2} \left[ \Sigma_{r,S}^{(\ell)} (p,m) \right] = 
-  \lim_{p^2 \to m^2} \left[ \Sigma_{r,V}^{(\ell)} (p,m) \right] \,,
\eeq
which is required to guarantee that the pole of the electron propagator remains at the location of 
physical on-shell electron mass.
The various orders explicitly read (again putting $\xi = 0$ for simplicity beyond one loop)
\begin{align}
\Sigma_{r,V}^{(1)}(p,\mr)|_{p^2=\mr^2}  &= 
\frac{\xi}{4 \epsilon}+\frac\xi2 + \mathcal{O}(\epsilon)
\,,
\\
\Sigma_{r,V}^{(2)}(p,\mr)|_{p^2=\mr^2} &=
-\frac{7}{64 \epsilon }+\frac{91}{128}-\frac{\pi ^2}{12}+\frac{1}{4} \pi ^2 \log
   (2)-\frac{3 \zeta (3)}{8} + \mathcal{O}(\epsilon)
   \,,
\\
\nonumber\Sigma_{r,V}^{(3)}(p,\mr)|_{p^2=\mr^2} &=
\frac{7}{144 \epsilon ^2}-\left( \frac{1517}{3456}+\frac{\pi ^2}{18}-\frac{1}{6} \pi ^2 \log (2)+\frac{\zeta (3)}{4}\right)\frac1\epsilon    \\
&  -\frac{20321}{20736}+\frac{9161 \pi ^2}{25920}-\frac{2371 \pi ^4}{17280}-\frac{5}{24} \pi
   ^2 \log (2)+\frac{8}{9} \pi ^2 \log ^2(2)  \\
\nonumber   & +\frac{25 \log ^4(2)}{144}+\frac{55}{72}  \zeta (3)+\frac{25 }{6}\text{Li}_4\!\left(\tfrac{1}{2}\right)+\frac{5 }{16}\zeta (5)    + \mathcal{O}(\epsilon)
   \,,
\end{align}
which are finite in the limit $x \to 1$.
We stress here that, of course, if one expands these formulas one order higher in $(1-x)$,
one can also verify that the residue at the pole is correctly fixed to be equal to one,
as a consequence of the wave function renormalization.


\subsection{Connection to modular forms}
\label{sec:modular}

We had already mentioned in section~\ref{sec:epsbasis} that the kernels $f_i$, with $i=10,\ldots,16$, that contain the solution $\varpi_0(x)$ of the Picard-Fuchs equation~\eqref{eq:pfell}, can be written in terms of modular forms. In this section, we elaborate more on this point, and we comment on which classes of transcendental functions enter the electron self-energy up to three loops and up to finite terms.
At one loop, only polylogarithms appear, and at two loops the electron self-energy can be expressed in terms of polylogarithms and iterated integrals of modular forms for the congruence subgroup $\Gamma_1(6)$~\cite{Honemann:2018mrb} (see below). As we will see, at three loops new classes of functions are required, albeit very similar to those encountered at lower loop orders.

We have already seen that the kernels $f_i$ with $i<10$ are dlog-forms. We therefore expect that this class of functions is closely related to multiple polyogarithms~\cite{Goncharov:1998kja,GoncharovMixedTate}. However, care is needed, because these kernels involve two square roots of 2 different quadratic polynomials, and it is not guaranteed that one can rationalize them simultaneously (see, e.g., refs.~\cite{Besier:2018jen,Besier:2019kco} for a review when or how one can rationalize square roots). It can indeed be checked that the electron self-energy contains in the finite term the iterated integrals $I(f_7,f_8)$ and $I(f_7,f_9)$, which depend on both square roots at the same time. In our case, the two square roots can be rationalized simultaneously, because they share the branch point at $x=1$. More precisely, if we perform for example the changes of variables
\beq\label{eq:rationalization}
x = \frac{y_1\,(3 - y_1)}{1 + y_1} \qquad \textrm{and}\qquad y_1 = \frac{6(1 + y_2)}{1+3 y_2^2}\,,
\eeq
then both square roots are rationalized:
\beq\bsp
\sqrt{(1-x)(9-x)} &\,= \frac{(1-y_1)(3+y_1)}{1+y_1}\\
&\, = \frac{3 \,(5+6 y_2-3 y_2^2) \,(3 y_2^2+2 y_2+3)}{(3 y_2^2+1) \,(3 y_2^2+6 y_2+7)}\,,\\
\sqrt{(1-x)(3+x)} &\,= \frac{(1-y_1)}{1+y_1}\,\sqrt{y_1^2-6y_1-3} \\
&\,=i \sqrt{3}\,\frac{ 9 y_2^4-54 y_2^2-24 y_2+5}{(3 y_2^2+1)\, (3 y_2^2+6 y_2+7)}\,.
\esp\eeq
It follows that all iterated integrals that involve only the kernels $\omega_i$ with $i<10$ can be expressed in terms of polylogarithms. Unlike at one and two loops, where the arguments of the polylogarithms were rational functions of the kinematic variable $x$, at three loops the polylogarithms have an algebraic dependence on $x$.

Let us now turn to the iterated integrals that depend on the kernels $f_i$ with $i\ge 10$. These kernels involve the solution $\varpi_0(x)$ of the Picard-Fuchs equation~\eqref{eq:pfell}, which can be expressed in terms of complete elliptic integrals and algebraic functions (cf.~eq.~\eqref{eq:pers}). As already mentioned, these kernels are closely related to modular forms. This is most easily seen after changing variables from $x$ to $\tau$ according to eq.~\eqref{eq:tau}. The inverse relation, which expresses $x$ as a function of $\tau$, is
\beq\label{eq:x_to_tau}
x = t(\tau) = 9\,\frac{\eta (\tau )^4\, \eta (6 \tau )^8}{\eta (2 \tau )^8\, \eta (3 \tau )^4}\,,
\eeq
where $\eta(\tau)$ is the Dedekind $\eta$-function,
\beq
\eta(\tau) = q^{\tfrac{1}{24}}\prod_{n=1}^\infty(1-q^n)\,,\qquad q=e^{2\pi i\tau}\,.
\eeq
The function $t(\tau)$ is a Hauptmodul for the congruence subgroup $\Gamma_1(6)$~\cite{Maier}, with
\beq
\Gamma_1(N) = \left\{\left(\begin{smallmatrix} a&b\\c&d\end{smallmatrix}\right)\in\textrm{SL}(2,\mathbb{Z}):a,d = 1\!\!\!\!\mod N \textrm{~and~} c = 0\!\!\!\!\mod N\right\}\,.
\eeq
In particular, $t(\tau)$ is invariant under the action of $\Gamma_1(6)$ by M\"obius transformations,
\beq
t\!\left(\tfrac{a\tau+b}{c\tau+d}\right) = t(\tau) \,,\qquad \left(\begin{smallmatrix} a&b\\c&d\end{smallmatrix}\right) \in \Gamma_1(6)\,.
\eeq
When expressed in terms of the variable $\tau$, the solution $\varpi_0$ of the Picard-Fuchs equation~\eqref{eq:pfell} becomes~\cite{Maier},
\beq\label{eq:per_to_psi}
\varpi_0(t(\tau)) = \psi(\tau) = \frac{\eta (2 \tau )^6 \eta (3 \tau )}{\eta (\tau )^3 \eta (6 \tau )^2}\,.
\eeq
It can be checked that $\psi(\tau)$ is a modular form (more specifically an Eisenstein series) of weight 1 for $\Gamma_1(6)$. In the following, we do not give a review of modular forms, but we refer to the literature (see, e.g., ref.~\cite{Zagier2008} for an introduction to modular forms). Here it suffices to say that $\psi(\tau)$ is holomorphic everywhere on the upper half-plane for $\tau$, including the cusps (i.e., the values of $\tau$ for which $t(\tau)\in \{0,1,9,\infty\}$, which are the singular points of the Picard-Fuchs equation~\eqref{eq:pfell}), and under $\Gamma_1(6)$ transformations it behaves like
\beq
\psi\!\left(\tfrac{a\tau+b}{c\tau+d}\right) = (c\tau+d)\,\psi(\tau) \,,\qquad \left(\begin{smallmatrix} a&b\\c&d\end{smallmatrix}\right) \in \Gamma_1(6)\,.
\eeq
Moreover, every modular form of weight $k\ge 1$ for $\Gamma_1(6)$ (with $k$ an integer) can be written in the form~\cite{Broedel:2018rwm}
\beq
\psi(\tau)^k\left[c_0+c_1 \,t(\tau) + \ldots + c_k\,t(\tau)^k\right]\,,\qquad c_i \in \mathbb{C}\,.
\eeq

Let us now discuss how modular forms arise from the kernels $f_i$ defined in section~\ref{sec:epsbasis}. The Jacobian of the change of variables from $x$ to $\tau$ is
\beq\label{eq:jac_tau}
\frac{\rd\! x}{\rd\! \tau} = \frac{i\pi}{8}\,\psi(\tau)^2\,t(\tau)\,(1-t(\tau))\,(9-t(\tau))\,.
\eeq
Using eqs.~\eqref{eq:x_to_tau},~\eqref{eq:per_to_psi} and~\eqref{eq:jac_tau}, we see that functions $\eta_j(\tau)$ defined in eq.~\eqref{eq:eta_f_def} define modular forms for $j\in\{3,5,6,10,11,12,14,15,16\}$. The weight of the modular forms is 2 for $j<10$, $0$ for $j=10$, $3$ for $j=11,12$ and 4 for $j\ge 14$. Note that these kernels contain the set of kernels that had appeared at two loops (cf.~eq.~\eqref{eq:lett2l}), in agreement with the fact that at two loops only iterated integrals of modular forms for $\Gamma_1(6)$ appear in the electron self-energy~\cite{Honemann:2018mrb}.

Let us now discuss the kernels $f_j$, with $j\in\{1,2,4\}$. While these kernels are dlog forms, they can appear in iterated integrals together with the kernels with $i\ge 10$, which are not dlog forms.\footnote{It is easy to see that in our case it is not possible to use, e.g., shuffle identities to decouple the dlog kernels from those that give rise to modular forms.} We can change variables from $x$ to $\tau$ for those integrals, and we see that for $i\le 3$, we can write
\beq\label{eq:mero_form}
\eta_j(\tau) = \psi(\tau)^2\left[\frac{c_{j,-1}}{t(\tau) - t_j} + c_{j,0} + c_{j,1}\,t(\tau) + c_{j,2}\,t(\tau)^2\right]\,, 
\eeq
with $c_{m,j} \in \mathbb{Q}$ and $c_{-1,j}\neq 0$, and $(t_1,t_2,t_4) = (-3,-1,2)$. The terms that are polynomial in $t(\tau)$ are modular forms of weight 2 for $\Gamma_1(6)$. The remaining term has a pole at $t(\tau_j) = t_j$. The poles in the upper half-plane are at points in the $\Gamma_1(6)$-orbit of 
\beq\bsp
\tau_{-3} &\, = \tfrac{1}{2}+i\,\tfrac{\sqrt{3}}{6}\,,\\
\tau_{-1} &\, = -\left(2 + i\,\tfrac{\K\left(1-\lambda_{-1}\right)}{\K(\lambda_{-1})}\right)^{-1}\,, \quad \lambda_{-1} = -\tfrac{3}{5}-i\tfrac{4}{5}\,,\\
\tau_{2} &\, = -\left(3 + i\,\tfrac{\K\left(1-\lambda_{2}\right)}{\K(\lambda_{2})}\right)^{-1}\,, \quad \lambda_{2} = \tfrac{1}{2}-i\tfrac{11}{16\sqrt{2}}\,.
\esp\eeq
Note that $\tau_{-3}$ is algebraic~\cite{Bonisch:2024nru}, while for $\tau_{-1}$ and $\tau_{2}$ we were not able to find an algebraic expression. Since the functions in eq.~\eqref{eq:mero_form} have a pole at values of $\tau$ with $\textrm{Im}(\tau)>0$, they are not holomorphic, and thus they do not define (holomorphic) modular forms. Instead, they are meromorphic modular forms of weight 2 for $\Gamma_1(6)$, whose iterated integrals have been studied in ref.~\cite{matthes_mero,Broedel:2021zij}.

Next, let us discuss the kernels $f_j$, or equivalently the functions $\eta_j(\tau)$, for $j\in\{8,9,13\}$. These kernels involve the square root $\sqrt{(1-x)(9-x)}$. The branch points of the square root coincide with the singular points $x=1,9$ of the Picard-Fuchs equation~\eqref{eq:pfell}, or equivalently the branch cut extends between two of the cusps of the modular curve for $\Gamma_1(6)$. Rationalizing this square root is equivalent to increasing the level of the modular forms. In this case we find that if we let 
\beq
y_1 = u(\tau) = 3\,\frac{ \eta (2 \tau )^2 \eta (12 \tau )^4}{\eta (4 \tau )^4 \eta (6 \tau )^2}\,,
\eeq
with $y_1$ defined in eq.~\eqref{eq:rationalization}, then $u(\tau)$ is a Hauptmodul for the congruence subgroup $\Gamma_1(12)$, and we have the relation (cf.~eq.~\eqref{eq:rationalization}):
\beq
t(\tau) = \frac{u(\tau)\,(3 - u(\tau))}{1 + u(\tau)}\,.
\eeq
Correspondingly, we find that the functions $\eta_j(\tau)$, for $j\in\{8,9,13\}$ are modular forms for $\Gamma_1(12)$ of weight 2 (for $j=8,9$) and 3 (for $j=13$). Hence, unlike at two loops, at three loops  we also encounter iterated integrals of modular forms for $\Gamma_1(12)$. Note that, since $\Gamma_1(12)\subset\Gamma_1(6)$, all modular forms for $\Gamma_1(6)$ are also modular forms for $\Gamma_1(12)$, but not vice versa.

Finally, let us discuss the kernel $f_{7}$, which involves the square root $\sqrt{(1-x)(3+x)}$. This kernel enters the finite part of the electron self-energy in iterated integrals, together with other kernels that are not dlog-forms. We can again change variables to $\tau$, but this time, since $x=-3$ is not a singular point of the Picard-Fuchs equation~\eqref{eq:pfell}, we are not able to interpret $\eta_7(\tau)$ as a modular form. In fact, $\eta_7(\tau)$ has branch cuts in the upper half plane starting from a point in the $\Gamma_1(6)$-orbit of $\tau_{-3}$ and ending at the cusp corresponding to $x=1$. We thus conclude that at three loops, a genuinely new class of iterated integrals arise that transcends the class of special functions (polylogarithms and iterated integrals of (meromorphic) modular forms) that have been studied in the literature.

\section{Analytic continuation and global solution}
\label{sec:global}

In this section, we describe how to obtain series solutions valid over the whole phase space, 
which allow us to evaluate $\Sigma^{(3)}_{N}$ (with $N=S,V$) 
for all values of the variable $x$ with $x\in \mathbb{R}+i 0$, where the positive small imaginary part is dictated by Feynman's prescription.\footnote{Clearly, the sign of the $i \epsilon$ is relevant only above threshold, where an explicit imaginary part is generated.} In order to achieve this, we follow the strategy outlined in section~\ref{sec:iterint}. Specifically, we solve the system of differential equations for the master integrals as series around each singular point $x_0 \in \{-3,-1,0,1,2,9,\infty\}$. We then use these results to derive the corresponding series expansions for $\Sigma^{(3)}_{N}$. 
As it is well known, the radius of convergence of a certain series expansion centered at 
any given $x_0$ is limited by the distance between $x_0$ and the closest singularity. It is important to stress here that
not all singularities of the master integrals survive in the expressions for $\Sigma^{(3)}_{N}$, which in turn means that
the convergence of the series for $\Sigma^{(3)}_{N}$ might be guaranteed on a larger interval compared to the individual master integrals.
We will ignore this issue here, and focus on obtaining valid series representations for all master integrals separately.

We can cover the full physically relevant parameter space $\mathbb{R}+i 0$ by local series expansions. For each patch in the parameter space located around a singular point of the differential equation, we obtain the boundary constants by matching a general solution in the form of eq.~(\ref{eq:expJ}) against numerical results obtained via \texttt{AMFlow}. Since we are aiming here at a purely numerical solution, we do not attempt any analytic fit (e.g. via PSLQ algorithm) of those boundary values, but we rather just store them as floats with a given precision. The only exceptions are $x_0=0$ and $x_0=1$ where all boundary constants have been derived analytically, as described in the previous sections\footnote{Some of the 
boundary constants at $x_0=0$ were also cross-checked with the results of ref.~\cite{Kniehl:2017ikj}.}.

Close to each of the singularities we redefine our master integrals in such a way that the arguments of the roots are positive definite in the region considered, i.e., we choose
\begin{equation}
    \begin{aligned}
        &\{\sqrt{(1-x)(9-x)},\,\sqrt{(3+x)(1-x)})\}\,, &&\qquad \qquad \text{for $-3<x<1$}\, ,\\ 
        &\{\sqrt{(x-1)(9-x)},\, \sqrt{(3+x)(x-1)})\}\,, &&\qquad \qquad  \text{for $1<x<9$}\, ,\\ 
        &\{\sqrt{(x-1)(x-9)},\,\sqrt{(3+x)(x-1)}\}\,, &&\qquad \qquad \text{for $x>9$ and $x<-3$}\, .
    \end{aligned}
\end{equation}
For $x_0 \in \{0,1,9,\infty\}$ explicit expressions for $\varpi_0$ were given in section \ref{sec:def}. The remaining 
singular points, namely, $x_0 \in \{-3,-1,2\}$, are regular points of the Picard-Fuchs equation~(\ref{eq:pfell}), and 
a power series solution is straightforward to obtain.

When crossing a given singular point $x_0$ of the differential equations, we perform the analytic continuation on
the logarithms  as follows
\begin{equation}
\begin{split}
    &\log(x_0-x)\,,       \qquad \qquad  \text{for } x<x_0\,,        \\
    &\log(x-x_0) -i\pi\,,  \qquad \text{for } x>x_0\,,     
\end{split}
\end{equation}
while for the square root we have
\begin{equation}
\begin{split}
      &\sqrt{x_0-x}\,,    \qquad \qquad \text{for } x<x_0\,,        \\
    -i&\sqrt{x-x_0}\,,  \qquad \qquad \text{for } x>x_0\,.     
\end{split}
\end{equation}

This guarantees that, in each region, the arguments of the logarithms and square roots are real and positive. 
In both cases, the signs of the imaginary parts are fixed using $x \to x + i \epsilon$.

To cover  the full parameter space, we have to pay special attention to the interval $(-9,-5)$; 
this interval is covered neither by the expansion around $x_0=-3$ nor $x_0=\infty$. 
We can proceed in two ways. The standard approach would consist in performing 
an additional series expansion around a non-singular point inside the interval $(-9,-5)$, 
in order to bridge between the two intervals.
Alternatively, we can rely on a Möbius transformation, as introduced in ref.~\cite{Lee:2017qql} 
and applied also in refs.~\cite{Hidding:2020ytt,Fael:2022rgm}. In our case, we used 
\begin{equation}
    x(z) = \frac{z-3}{z+1}, \qquad  \text{with inverse:\quad} z(x)=-\frac{x+3}{x-1}  \, ,
\label{eq:Mobius_transformation}
\end{equation}
which maps the points $\{\infty,-3,-1\}$ (in the original variable $x$) to $\{-1,0,+1\}$ (in the variable $z$), respectively. 
This means that the original interval $(\infty,-1)$ is mapped to $(-1,1)$. 
All the other singular points $x_0$ are mapped outside the unit disk. 
Therefore, thanks to eq.~\eqref{eq:Mobius_transformation}, starting from a power series centered at $x_0=-3$, 
we can obtain a power series in the variable $z$, which is convergent inside the unit disk. This allows us to cover also the points corresponding to $(-9,-5)$ in the variable $x$. 

Let us now comment on the numerical performances of the series expansions so obtained. 
Since we have computed the boundary constants independently at each singularity through \texttt{AMFlow}, 
we can use the difference of two series expansions in a common region of convergence to estimate 
the accuracy of these expansions. For example, let us focus for simplicity on $\Sigma^{(3)}_{S,0}$
and consider the point $x=1/2$. We also assume for definiteness that $\xi=0$. 
To evaluate $\Sigma^{(3)}_{S,0}$ at $x=1/2$, 
we can use the series expansions derived around $x=0$ or $x=1$. 
Since both expansion points have the same distance to $x=1/2$, the difference will give a conservative 
estimate of the precision of the series expansion of $\Sigma^{(3)}_{N}$ in the interval $(0,1)$, 
when we take for other points in $(0,1)$ the series expansions derived at the closer singularity. 
Taking $100$ terms of the series solutions into account, we find
\begin{equation}
\begin{aligned}
    &\Sigma^{(3)}_{S,0} \left(1/2 \right)=\num{-4.65474313859850740532525589187}, \qquad\text{($x_0=0$)}, \\
    &\Sigma^{(3)}_{S,0}\left( 1/2 \right)=\num{-4.65474313859850740532525589274}, \qquad\text{($x_0=1$)},
\end{aligned}
\label{num12}
\end{equation}
which agree up to 27 digits. To see how fast the two series expansions converge, 
 in table~\ref{tab:relerrorx12} we showcase the relative error of $\Sigma^{(3)}_{S,0}$ compared to eq.~\eqref{num12}, when we truncate 
 the series expansion at a lower order.
\begin{table}[h]
    \centering
    \begin{tabular}{|c||c|c|}
    \hline
     \text{Truncation Order}    & Rel. Error Series $x_0=0$    & Rel. Error Series $x_0=1$ \\
\hline
 10 & $0.6 \times 10^{-2}$  & $0.2 \times 10^{-1}$   \\
 30 & $8.9 \times 10^{-9}$  & $9.6 \times 10^{-8}$   \\
 50 & $1.0 \times 10^{-14}$ & $1.8 \times 10^{-13}$   \\
 90 & $1.1 \times 10^{-26}$ & $2.1 \times 10^{-25}$   \\
 95 & $3.5 \times 10^{-28}$ & $6.0 \times 10^{-27}$   \\
 \hline
    \end{tabular}
    \caption{Relative error of partial sums derived around $x_0=0,1$ with different truncation orders against the result at $x=1/2$ computed with $100$ terms.}
    \label{tab:relerrorx12}
\end{table}
A similar analysis can be also made at $x=3/2$, which is above the $p^2 = m^2$ threshold and 
the self-energy develops an imaginary part. We find
\begin{equation}
\begin{aligned}
    \Sigma^{(3)}_{S,0}(3/2) = &-      \num{39.0874347745994094234625921809} \\
    & +  \num{44.6006079875072823586914700677} \, i, \qquad (x_0=1), \\
    \Sigma^{(3)}_{S,0}(3/2) = &-\num{39.0874347745994094234625922288} \\
    &+\num{44.6006079875072823586914700279}i, \qquad (x_0=2),
\end{aligned}
\label{num32}
\end{equation}
which also agree to 27 digits. 
Again, we show in table~\ref{tab:relerrorx32} the convergence of the series expansions.
\begin{table}[h]
    \centering
    \begin{tabular}{|c||c|c|}
    \hline
     \text{Truncation Order}    & Rel. Error Series $x_0=1$    & Rel. Error Series $x_0=2$ \\
\hline
 10 & $0.4 \times 10^{-2}$  & $0.8 \times 10^{-3}$   \\
 30 & $9.0 \times 10^{-9}$  & $1.0 \times 10^{-9}$   \\
 50 & $7.4 \times 10^{-14}$ & $1.1 \times 10^{-15}$   \\
 90 & $7.4 \times 10^{-25}$ & $1.1 \times 10^{-27}$   \\
 95 & $2.9 \times 10^{-26}$ & $3.0 \times 10^{-29}$   \\
 \hline
    \end{tabular}
    \caption{Relative error of partial sums derived around $x_0=1,2$ with different truncation orders against the result at $x=3/2$ computed with $100$ terms.}
    \label{tab:relerrorx32}
\end{table}

\begin{figure}[H]
    \centering
    \includegraphics[width=0.6\linewidth]{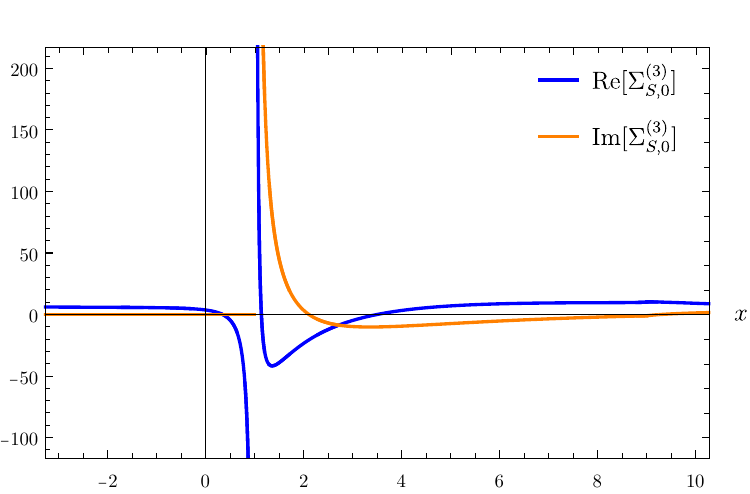}
    \caption{Real and imaginary part of $\Sigma^{(3)}_{S,0}$ (for $\xi=0$).}
    \label{fig:Scalar_global}
\end{figure}

\begin{figure}[H]
    \centering
    \includegraphics[width=0.6\linewidth]{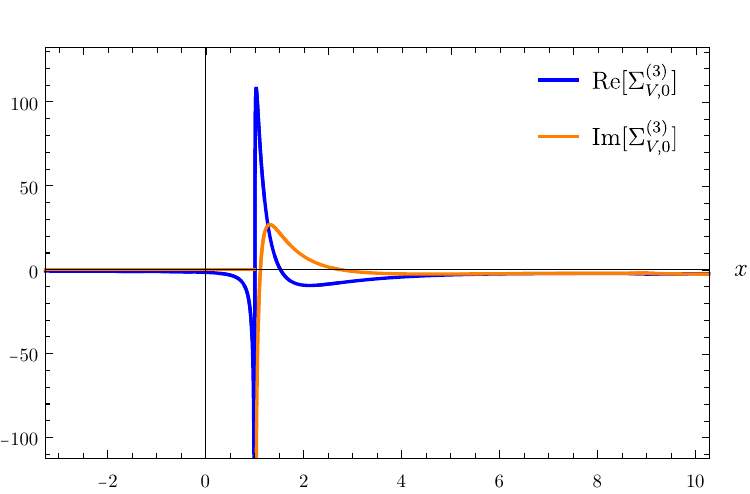}
    \caption{Real and imaginary part of $\Sigma^{(3)}_{V,0}$ (for $\xi=0$).}
    \label{fig:Vector_global}
\end{figure}

\begin{figure}[h]
    \centering
    \includegraphics[width=0.6\linewidth]{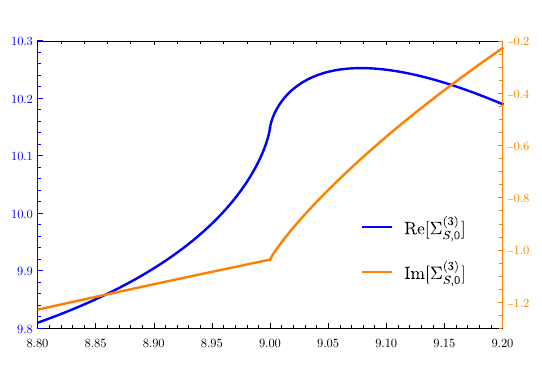}
    \caption{Real and imaginary part of $\Sigma^{(3)}_{S,0}$ close to $x_0=9$ (for $\xi=0$). Notice different scales for different quantities in ordinate.}
    \label{fig:x_0=9_scalar}
\end{figure}

\begin{figure}[h]
    \centering
    \includegraphics[width=0.6\linewidth]{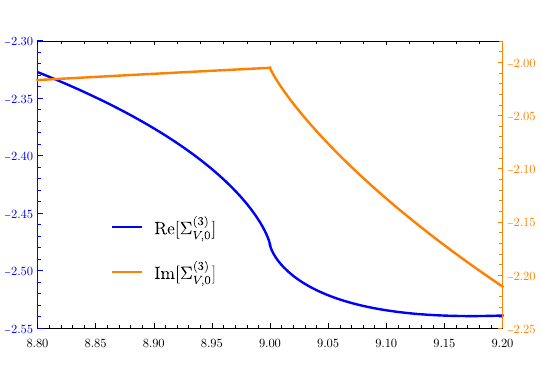}
    \caption{Real and imaginary part of $\Sigma^{(3)}_{V,0}$ close to $x_0=9$ (for $\xi=0$). Notice different scales for different quantities in ordinate.}
    \label{fig:x_0=9_vector}
\end{figure}

Having discussed the numerical error of our series expansions, 
we provide here several plots showing $\Sigma^{(3)}_{N}$ for $N=S,V$ assuming $\xi=0$, in fig.~\ref{fig:Scalar_global} and fig.~\ref{fig:Vector_global}, respectively.\\
From the plots in figs.~\ref{fig:Scalar_global} and~\ref{fig:Vector_global}, we can easily see the expected discontinuity generated by the threshold singularity at $x_0=1$. Notice the vanishing imaginary part below $x_0=1$. At the point $x_0=9$, the functions are continuous but not differentiable, see figs.~\ref{fig:x_0=9_scalar} and~\ref{fig:x_0=9_vector}.

\subsection{Accelerated series expansions}

It is a well known fact that one can apply suitable transformations in order to 
improve the rate of convergence of a certain class
of series expansions. When dealing with generalized series with logarithmic singularities of the type that typically appear in the
evaluation of classical or multiple polylogarithms, 
a popular choice is given by so-called Bernoulli-like variables, first introduced in ref.~\cite{tHooft:1978jhc}.
These have since been  exploited in most numerical routines for the numerical evaluation of multiple polylogarithms, see,~e.g.,~refs.~\cite{Gehrmann:2001jv,Vollinga:2004sn,Buehler:2011ev}. 
Here, we examine their use to accelerate the series expansions for the Feynman integrals encountered in our calculation,
which are related to a more general elliptic geometry. 
The effectiveness of using Bernoulli-like variables beyond polylogarithms has already been
 demonstrated in the case of the two-loop equal mass sunrise graph~\cite{Pozzorini:2005ff}.
In fact, irrespective of the geometry involved, 
close to each regular singular point, Feynman integrals
are always expected to be expressible as generalized series expansion of the form given in eq.~\eqref{eq:expJ}.
Bernoulli-like variables are related to the logarithmic behavior of the corresponding series 
at the singular point considered, and we expect therefore that they could help accelerate the convergence 
whenever these types of series are considered.

Let us showcase the effect of this transformation, reconsidering 
the series expansion for $\Sigma^{(3)}_{S,0}$ around $x_0=0$. The variable we consider is defined by
\begin{equation}
x(z) = 1-e^{-z}, \qquad \text{with inverse:} \qquad z(x)=- \log(1-x)\,.
\label{eq:Bernoulli_transformation}
\end{equation}
We can express $\Sigma^{(3)}_{S,0}$ as a function of $z$ thanks to eq.~(\ref{eq:Bernoulli_transformation}) and study the convergence of partial sums at the point $z=\log(2)$ (corresponding to $x=1/2$ in the original variables). The result is reported in table~\ref{tab:values_x0_Bernoulli}, where one can appreciate a much faster convergence compared to table~\ref{tab:relerrorx12}.
In particular, with only $36$ terms the series expansion in the Bernoulli-like variable 
achieves a precision of $30$ digits, which is beyond what was obtained through the original series including  $100$ terms.
\begin{table}[H]
    \centering
    \begin{tabular}{|c||c|}
    \hline
     \text{Truncation Order}    & Rel. Error Series $x_0=0$ \\
\hline
 10 & $2.3 \times 10^{-8}$ \\
 20 & $5.3 \times 10^{-19}$ \\
 30 & $3.4 \times 10^{-26}$ \\
 40 & $1.7 \times 10^{-33}$ \\
 \hline
    \end{tabular}
    \caption{Convergence of the series expansion solution $\Sigma^{(3)}_{S,0}$ in the Bernoulli-like variable around $x_0=0$, evaluated $x=1/2$.
    We provide the result for different truncation orders to showcase the series convergence.
    }
    \label{tab:values_x0_Bernoulli}
\end{table}
Similarly, one can attempt to obtain accelerated series expansions close to the other singular points, using the appropriate variable transformation. As it turns out, changing to a Bernoulli-like variable works very well at accelerating the convergence of the corresponding series close to singular points. How much the convergence is accelerated seems to strongly depend on the specific singular point considered. In our case, we observe that the convergence improvement is the smallest close to the on-shell limit, i.e. $p^2=m^2$, $x=1$. Studying the origin of this behavior and the general applicability of Bernoulli-like accelerated series expansions goes beyond the scope of the current paper. We postpone this investigation to a separate dedicated publication.

Making use of accelerated expansions, in the 
ancillary files we provide results in the form of a Mathematica routine that can evaluate the electron self-energy to three loops in all 
regions of the parameter space with a fixed precision of at least $16$ correct digits. For this routine, we have derived the series expansions in all patches up to 50 terms in Bernoulli-like variables, including the full $\xi$-dependence. These series are attached to the ancillary file.


\section{Conclusions}
\label{sec:conc}
In this paper, we have addressed the analytic and numerical evaluation of the three-loop corrections
to the electron self-energy in QED. Starting from its Feynman diagram representation, we have used
integration-by-parts identities to express the relevant scalar functions in terms of master integrals
and derived a system of differential equations for the latter.
We have shown that all master integrals are either of polylogarithmic type or, at most, related to
the elliptic curve of the two-loop sunrise graph.
Leveraging new developments in the theory of differential equations for these type of geometries,
we have derived an $\epsilon$-factorized system of differential equations, whose connection
matrix is expressed in terms of independent differential forms.
We have then used this basis to obtain, on the one hand, formal analytic solutions in terms of linearly
independent iterated integrals, and on the other, series expansion solutions valid close to all singular points.
In the former representation, we have shown that all iterated integrals involving new differential forms
that appear for the first time at three loops, are explicitly confined in the finite remainder of the self-energy.
This representation allowed us also to perform analytically UV renormalization, since all poles are naturally
expressed only in terms of differential forms that appear at lower loops.
In discussing its analytic properties, we have shown that a 
new class integrals, beyond that of iterated integrals over 
meromorphic modular forms, appears in this calculation. To the best of our knowledge, these integrals have not been considered in the mathematical literature.

The series expansion representation for the solution has instead been used to obtain fast and reliable numerics for the self-energy across the whole parameter space.
While we have typically produced series representation using numerical boundary conditions in the various
regions of interest, we have studied two points in more detail, namely $p^2=0$ and $p^2=m^2$.
For what concerns $p^2=0$, we have shown that the self-energy can be expressed in terms of multiple polylogarithms evaluated
at sixth roots of unity. We have provided an alternative basis for these numbers up to weight four, which allows us to maximally
express all constants in terms of classical polylogarithms.
We have then dedicated special attention to derive an $\epsilon$-resummed series representation 
close to the point $p^2 = m^2$, which allowed us to obtain at once both the mass and the wave function renormalization
constants in the on-shell scheme. Here we have shown that, as expected, our results up to finite order in $\epsilon$ 
can be written in terms of multiple zeta values
and classical polylogarithms evaluated at $x=\frac12$.
Finally, we have studied the effect of introducing Bernoulli-like variables  
on the convergence of the series expansion and observed that, while they substantially 
accelerate the convergence of the series expansions at most regular singular points, they do not seem to work equally well everywhere. The least improvement is observed close to the on-shell limit $p^2=m^2$.

The results obtained in this paper are interesting as they demonstrate, in a rather simple set up, how many
techniques recently developed for the calculation of master integrals and amplitudes are now mature enough to be utilized beyond the
real of multiple polylogarithms. We have demonstrated this both analytically and numerically. 
Analytically, we have seen how new ideas for the derivation of
$\epsilon$-factorized bases of master integrals and to treat the ensuing iterated integrals proved
successful to handle the mathematical structures encountered in this calculation, which involve the iterated
integration over admixtures of elliptic and logarithmic kernels. 
Numerically, we have proved the efficacy of 
standard techniques based on series expansions, and also studied the effect
of using accelerated series expansions based on
Bernoulli-like variables, typically used in the context of multiple polylogarithms, to treat also realistic elliptic Feynman diagrams.

\acknowledgments
We thank Federico Buccioni, Fabrizio Caola, Felix Forner, Nikolaos Syrrakos and 
Fabian Wagner for inputs and discussions and for collaboration on related projects.
LT is grateful to Ettore Remiddi for private communication about the master integrals
appearing in the calculation of the three-loop QED corrections to the electron $g$-2 and
on the use of Bernoulli-like variables to accelerate the convergence of series expansions beyond polylogs. FG would like to thank Hjalte Frellesvig for sharing with him a Mathematica package for the Baikov representation, and for related discussions.

This work was supported in part by the Excellence Cluster ORIGINS funded by the
Deutsche Forschungsgemeinschaft (DFG, German Research Foundation) under Germany’s
Excellence Strategy – EXC-2094-390783311,
the Excellence Cluster "Precision Physics, Fundamental Interactions, and Structure of
Matter" (Grant No.~EXC-2118-390831469) 
and in part by the European Research Council (ERC) under the European Union’s research and innovation programme grant agreements 949279
(ERC Starting Grant HighPHun) and
101043686 (ERC Consolidator Grant LoCoMotive). Views and opinions expressed are
however those of the author(s) only and do not necessarily reflect those of the European
Union or the European Research Council. Neither the European Union nor the granting
authority can be held responsible for them.

\appendix

\section{The canonical basis}
\label{app:can}

In this appendix we give a basis of master integrals, which puts the differential equation
into an $\eps$-factorized form. We set $m=1$ and use the dimensionless variable $x$.
The $\eps$-factorized basis introduces two square roots, which we denote by
\begin{eqnarray}
    r_1 \; = \; \sqrt{\left(1-x\right)\left(9-x\right)},
    & &
    r_2 \; = \; \sqrt{\left(1-x\right)\left(3+x\right)}.
\end{eqnarray}
$\myperiod_0$ denotes a solution of the Picard-Fuchs equation~(\ref{eq:pfell}),
${\bf D}^-$ denotes the dimensional-shift operator, which lowers the space-time dimension by two units.
The Jacobian $J$ is given by
\begin{eqnarray}
    J & = & \frac{1}{9} x \left(x-1\right)\left(x-9\right) \myperiod_0^2 \,.
\end{eqnarray}
The basis reads:
\begin{alignat}{2}
 \mbox{Sector 56:} \;\;\;\; &
 J_{1}
 & = \;\; & 
 - \eps^3 
 \; {\bf D}^- I_{000111000},
 \nonumber \\
 \mbox{Sector 57:} \;\;\;\; &
 J_{2}
 & = \;\; & 
 \eps^3 
 \left(1-x\right)
 \; {\bf D}^- I_{100111000},
 \nonumber \\
 \mbox{Sector 85:} \;\;\;\; &
 J_{3}
 & = \;\; & 
 \eps^3 
 \left(1-x\right)
 \; {\bf D}^- I_{101010100},
 \nonumber \\
 &
 J_{4}
 & = \;\; & 
 - \eps^3 
 \; {\bf D}^- I_{101\left(-1\right)10100},
 \nonumber \\
 \mbox{Sector 120:} \;\;\;\; &
 J_{5}
 & = \;\; & 
 \eps^3 
 \frac{1}{\myperiod_0}
 \; {\bf D}^- I_{000111100},
 \nonumber \\
 &
 J_{6}
 & = \;\; & 
 \frac{1}{\eps} J \frac{d}{dx} J_5
 + \frac{1}{18} \myperiod_0^2 \left(3x^2-10x-9\right) J_5,
 \nonumber \\
 \mbox{Sector 195:} \;\;\;\; &
 J_{7}
 & = \;\; & 
 \eps^2 \left(1+4\eps\right)
 \; {\bf D}^- I_{110000110},
 \nonumber \\
 \mbox{Sector 202:} \;\;\;\; &
 J_{8}
 & = \;\; & 
 \eps^3 
 \frac{1}{\myperiod_0}
 \; {\bf D}^- I_{010100110},
 \nonumber \\
 &
 J_{9}
 & = \;\; & 
 - \eps^3 
 \left[ {\bf D}^- I_{0101\left(-1\right)0110} + \left(1-x\right) \; {\bf D}^- I_{010100110} \right],
 \nonumber \\
 &
 J_{10}
 & = \;\; & 
 \frac{1}{\eps} J \frac{d}{dx} J_8
 + \frac{1}{18} \myperiod_0^2 \left(7x^2-30x-9\right) J_8,
 \nonumber \\
 \mbox{Sector 216:} \;\;\;\; &
 J_{11}
 & = \;\; & 
 \eps^3 
 \; {\bf D}^- I_{000110110},
 \nonumber \\
 \mbox{Sector 263:} \;\;\;\; &
 J_{12}
 & = \;\; & 
 \eps^3 
 \left(1-x\right)
 \; {\bf D}^- I_{111000001},
 \nonumber \\
 &
 J_{13}
 & = \;\; & 
 - \eps^3 
 \; {\bf D}^- I_{1110000\left(-1\right)1},
 \nonumber \\
 \mbox{Sector 59:} \;\;\;\; &
 J_{14}
 & = \;\; & 
 - \eps^3 
 \left(1-x\right)^2
 \; {\bf D}^- I_{110111000},
 \nonumber \\
 \mbox{Sector 87:} \;\;\;\; &
 J_{15}
 & = \;\; & 
 - \eps^3 
 \left(1-x\right)^2
 \; {\bf D}^- I_{111010100},
 \nonumber \\
 &
 J_{16}
 & = \;\; & 
 \eps^3 
 \left(1-x\right)
 \; {\bf D}^- I_{111\left(-1\right)10100},
 \nonumber \\
 \mbox{Sector 122:} \;\;\;\; &
 J_{17}
 & = \;\; & 
 - \eps^3 
 \frac{\left(1-x\right)}{\myperiod_0}
 \; {\bf D}^- I_{010111100},
 \nonumber \\
%
%
 &
 J_{18}
 & = \;\; & 
 \frac{1}{\eps} J \frac{d}{dx} J_{17}
 + \frac{1}{18} \myperiod_0^2 \left(5x^2-26x-27\right) J_{17}
 - \frac{1}{9} \myperiod_0^2 \left(x-1\right)\left(x-9\right) J_{5},
 \nonumber \\
 \mbox{Sector 199:} \;\;\;\; &
 J_{19}
 & = \;\; & 
 \eps^3 
 \left(1-x\right)
 \; {\bf D}^- I_{11100\left(-1\right)110},
 \nonumber \\
 &
 J_{20}
 & = \;\; & 
 - \eps^3 
 \; {\bf D}^- I_{11100\left(-2\right)110},
 \nonumber \\
 \mbox{Sector 217:} \;\;\;\; &
 J_{21}
 & = \;\; & 
 - \eps^3 
 \left[ \left(1-x\right) \; {\bf D}^- I_{100110110} + {\bf D}^- I_{100010110} \right],
 \nonumber \\
 \mbox{Sector 248:} \;\;\;\; &
 J_{22}
 & = \;\; & 
 \eps^3 
 r_1
 \; {\bf D}^- I_{00\left(-1\right)111110},
 \nonumber \\
 &
 J_{23}
 & = \;\; & 
 - \eps^3 
 \left[ {\bf D}^- I_{00\left(-2\right)111110} - 4 \; {\bf D}^- I_{00\left(-1\right)111110}\right],
 \nonumber \\
 &
 J_{24}
 & = \;\; & 
 \eps^3 
 \left(1-x\right)
 \left[ {\bf D}^- I_{00\left(-1\right)111110} - 4 \; {\bf D}^- I_{000111110}\right],
 \nonumber \\
 \mbox{Sector 271:} \;\;\;\; &
 J_{25}
 & = \;\; & 
 \eps^3 
 \left(1-x\right)
 \; {\bf D}^- I_{111100\left(-1\right)01},
 \nonumber \\
 \mbox{Sector 341:} \;\;\;\; &
 J_{26}
 & = \;\; & 
 \eps^3 
 \frac{1}{\myperiod_0}
 \left[ {\bf D}^- I_{101\left(-1\right)10101} 
      + {\bf D}^- I_{10101\left(-1\right)101}
      + \left(1-x\right) {\bf D}^- I_{101010101}
 \right. \nonumber \\
 & & & \left.
      - {\bf D}^- I_{101010001}
 \right],
 \nonumber \\
 &
 J_{27}
 & = \;\; & 
 \eps^3
 \left\{
   \left[ - {\bf D}^- I_{1\left(-1\right)1\left(-1\right)10101} 
      - {\bf D}^- I_{1\left(-1\right)101\left(-1\right)101}
      - \left(1-x\right) {\bf D}^- I_{1\left(-1\right)1010101}
 \right. \right. \nonumber \\
 & & & \left. \left.
      + {\bf D}^- I_{1\left(-1\right)1010001}
 \right]
 + \frac{1}{3} \left(x+3\right)
 \left[ {\bf D}^- I_{101\left(-1\right)10101} 
      + {\bf D}^- I_{10101\left(-1\right)101}
 \right. \right. \nonumber \\
 & & & \left. \left.
      + \left(1-x\right) {\bf D}^- I_{101010101}
      - {\bf D}^- I_{101010001}
 \right]
 \right\},
 \nonumber \\
 &
 J_{28}
 & = \;\; & 
 \frac{1}{\eps} J \frac{d}{dx} J_{26}
 + \frac{1}{54} \myperiod_0^2 \left(13x^2-66x-27\right) J_{26},
 \nonumber \\
 \mbox{Sector 63:} \;\;\;\; &
 J_{29}
 & = \;\; & 
 \eps^3 
 \left(1-x\right)^3
 \; {\bf D}^- I_{111111000},
 \nonumber \\
 \mbox{Sector 125:} \;\;\;\; &
 J_{30}
 & = \;\; & 
 - \eps^4 \left(1-2\eps\right)
 x \; I_{101121100},
 \nonumber \\
 \mbox{Sector 207:} \;\;\;\; &
 J_{31}
 & = \;\; & 
 - \eps^4 \left(1-2\eps\right)
 x \; I_{111100120},
 \nonumber \\
 &
 J_{32}
 & = \;\; & 
 - \eps^4 \left(1-2\eps\right)
 x \; I_{121100110},
 \nonumber \\
 \mbox{Sector 222:} \;\;\;\; &
 J_{33}
 & = \;\; & 
 \eps^3 
 \left[ 
         - \left(1-x^2\right) \; {\bf D}^- I_{11011\left(-1\right)110} 
         - \left(1-x^2\right) \; {\bf D}^- I_{11011011\left(-1\right)} 
 \right. \nonumber \\
 & & & \left.
         + 2 x \left(1-x\right) \; {\bf D}^- I_{100110110}
         + 4 \left(1-x\right) \; {\bf D}^- I_{110111000}
         + 2 x \; {\bf D}^- I_{010100110}
 \right],
 \nonumber \\
 &
 J_{34}
 & = \;\; & 
 - \eps^4 \left(1-2\eps\right)
 x \; I_{110110210},
 \nonumber \\
 \mbox{Sector 249:} \;\;\;\; &
 J_{35}
 & = \;\; & 
 - \eps^4 \left(1-2\eps\right)
 x \; I_{100111120},
 \nonumber \\
 \mbox{Sector 287:} \;\;\;\; &
 J_{36}
 & = \;\; & 
 \eps^3 
 \left[ \left(1-x\right) 
        \left( 
              - {\bf D}^- I_{11111\left(-1\right)001}
              + {\bf D}^- I_{111110\left(-1\right)01}
              + {\bf D}^- I_{1111100\left(-1\right)1}
        \right)
 \right. \nonumber \\
 & & & \left.
       - \frac{1}{3} \left(x+3\right) {\bf D}^- I_{010100110}
 \right],
 \nonumber \\
 &
 J_{37}
 & = \;\; & 
 - \eps^4 \left(1-2\eps\right)
 x \; I_{112110001},
 \nonumber \\
 &
 J_{38}
 & = \;\; & 
 - \eps^4 \left(1-2\eps\right)
 x \; I_{111110002},
 \nonumber \\
 \mbox{Sector 349:} \;\;\;\; &
 J_{39}
 & = \;\; & 
 - \eps^4 \left(1-2\eps\right)
 x \; I_{101120101},
 \nonumber \\
 \mbox{Sector 455:} \;\;\;\; &
 J_{40}
 & = \;\; & 
 - \eps^4 \left(1-2\eps\right)
 x \; I_{112000111},
 \nonumber \\
 &
 J_{41}
 & = \;\; & 
 \frac{\left(1-x\right)}{\eps} \frac{d}{dx} J_{40}
 - \frac{1}{x} J_{40}
 + \frac{1}{2x} \left( J_{20} - J_{19} + J_{4} - J_{3} \right)
 + \frac{\left(x-9\right)}{6} \myperiod_0 J_{8}
 - \frac{1}{6x} J_{7},
 \nonumber \\
 \mbox{Sector 462:} \;\;\;\; &
 J_{42}
 & = \;\; & 
 - \eps^4 \left(1-2\eps\right)
 x \; I_{011100121},
 \nonumber \\
 &
 J_{43}
 & = \;\; & 
 \frac{1}{\eps} \left(1-x\right) \frac{d}{dx} J_{42}
 - \frac{1}{x} J_{42}
 - \frac{1}{4} \myperiod_0 \left(1-x\right) J_{26}
 + \frac{1}{2} \myperiod_0 \left(1-x\right) J_{5},
 \nonumber \\
 \mbox{Sector 127:} \;\;\;\; &
 J_{44}
 & = \;\; & 
 \eps^4 \left(1-2\eps\right)
 x \left(1-x\right)
 \left[ I_{121111100} + I_{111121100} \right],
 \nonumber \\
 \mbox{Sector 251:} \;\;\;\; &
 J_{45}
 & = \;\; & 
 - \eps^5 \left(1-2\eps\right)
 x \; I_{110111110},
 \nonumber \\
 \mbox{Sector 351:} \;\;\;\; &
 J_{46}
 & = \;\; & 
 - \eps^5 \left(1-2\eps\right)
 x \; I_{111110101},
 \nonumber \\
 &
 J_{47}
 & = \;\; & 
 \frac{1}{\eps} x \frac{d}{dx} J_{46}
 - \frac{1}{1-x} \left[2 J_{46} + 2 J_{39} - J_{38} \right],
 \nonumber \\
 \mbox{Sector 463:} \;\;\;\; &
 J_{48}
 & = \;\; & 
 - \eps^5 \left(1-2\eps\right)
 x \; I_{111100111},
 \nonumber \\
 \mbox{Sector 255:} \;\;\;\; &
 J_{49}
 & = \;\; & 
 \eps^5 \left(1-2\eps\right)
 x \left(1-x\right)
 \; I_{111111110},
 \nonumber \\
 \mbox{Sector 479:} \;\;\;\; &
 J_{50}
 & = \;\; & 
 \eps^5 \left(1-2\eps\right)
 x r_2
 \; I_{111110111},
 \nonumber \\
 &
 J_{51}
 & = \;\; & 
 \frac{1}{\eps} r_2 \frac{d}{dx} J_{50}
 + \frac{\left(5x^2+4x+3\right)r_2}{x\left(x-1\right)\left(x+3\right)} J_{50}.
\end{alignat}

\section{QED on-shell renormalization constants to three loops}
\label{app:renconsts}
In this appendix, we present explicit results for the renormalization
constants to three loops, which were first obtained in ref.~\cite{Melnikov:2000zc}.
We expand them in the electric coupling as follows
\begin{align}
    &Z_m = 1 + \sum_{\ell=1}^\infty \left( C_r(\epsilon) \frac{\alpha}{\pi} \right)^\ell  Z_m^{(\ell)}\,, \nonumber \\
    &Z_2 = 1 + \sum_{\ell=1}^\infty \left(  C_r(\epsilon) \frac{\alpha}{\pi} \right)^\ell  Z_2^{(\ell)}\,,
\end{align}
where $C_r(\epsilon)$ was defined in~\cref{eq:defCReps}.
Including higher orders $\epsilon$ (which are necessary for the full
renormalization of the electron propagator to three-loops) 
their coefficients read
\begin{align}
    Z_m^{(1)} = &-\frac{3}{4 \epsilon }-\frac{1}{1-2 \epsilon }\,, \nonumber \\
    Z_m^{(2)} = &+ \frac{13}{32 \epsilon ^2}    +\frac{73}{64 \epsilon}
    -\frac{3 \zeta (3)}{4}
   -\frac{23 \pi ^2}{48}+\frac{475}{128}+\frac{1}{2} \pi ^2 \log(2) 
    + \epsilon \left[
   \frac{2529}{256}    
   -\frac{245 \pi^2}{96}   \right. \nonumber \\
    & \left. 
   -\frac{47 \zeta (3)}{4}
   +4 \pi ^2 \log(2)
   +\frac{7 \pi ^4}{40} -\frac{\log ^4(2)}{2}
   -\pi ^2 \log ^2(2)
   -12 \text{Li}_4\!\left(\tfrac{1}{2}\right) 
   \right]\,, \\
    Z_m^{(3)} = &-\frac{221}{1152 \epsilon ^3}
    -\frac{5561}{6912 \epsilon ^2}
    +\frac{1}{\epsilon } \left[ \frac{13 \zeta(3)}{16}-\frac{154445}{41472}
    +\frac{391 \pi ^2}{576}-\frac{17}{24} \pi ^2 \log(2)\right] \nonumber \\   
    &-\frac{3489365}{248832}-\frac{5783 \pi ^2}{17280}
        +\frac{89}{36} \pi ^2 \log (2) + \frac{719 \zeta(3)}{72}
    -\frac{979 \pi^4}{4320}
    +\frac{23 \log^4(2)}{72} \nonumber \\
    &+\frac{65}{36} \pi ^2 \log ^2(2)
    +\frac{23 }{3} \text{Li}_4\!\left(\tfrac{1}{2}\right)
    -\frac{\pi ^2 \zeta (3)}{16}+\frac{5 \zeta (5)}{8}\,,
\end{align}

\begin{align}
    Z_2^{(1)} = &-\frac{3}{4 \epsilon }
    -\frac{1}{1-2 \epsilon }\,, \nonumber \\
    Z_2^{(2)} = &-\frac{3 \zeta (3)}{2}+\frac{17}{32
   \epsilon ^2}+\frac{229}{192 \epsilon }-\frac{55 \pi ^2}{48}+\frac{8453}{1152}+\pi ^2
   \log (2) + \epsilon  \left[\frac{86797}{6912}-\frac{419 \pi ^2}{96}
   \right. \nonumber \\
& \left.
-\frac{203 \zeta(3)}{8}
   +\frac{7 \pi ^4}{20}
   -\log ^4(2)
   -2 \pi^2 \log ^2(2)
   +\frac{31}{4} \pi ^2 \log (2) 
   -24 \text{Li}_4\!\left(\tfrac{1}{2}\right) 
   \right]\,,
 \\
    Z_2^{(3)} = &    
    -\frac{131}{384 \epsilon ^3}
    -\frac{2141}{2304 \epsilon ^2}
    +\frac{1}{\epsilon } \left[\frac{17 \zeta(3)}{8}
    -\frac{116489}{13824}
    +\frac{935 \pi ^2}{576}
    -\frac{17}{12} \pi ^2 \log(2)\right] \nonumber \\
    &   -\frac{2121361}{82944}
    -\frac{197731 \pi ^2}{51840}
    +\frac{1367}{144} \pi ^2 \log (2)
    +\frac{2803 \zeta(3)}{144}
    -\frac{383 \pi^4}{720}
    +\frac{3 \log^4(2)}{4} \nonumber \\
    &+\frac{23}{6} \pi ^2 \log ^2(2)
     + 18 \text{Li}_4\!\left(\tfrac{1}{2}\right)
     +\frac{\pi ^2 \zeta (3)}{8}-\frac{5 \zeta (5)}{16}\,.
\end{align}

\bibliographystyle{JHEP} 
\bibliography{biblio} 

\end{document}